\documentclass[aps,prb,twocolumn,showpacs,amsmath,amssymb,superscriptaddress,bibnotes,longbibliography]{revtex4-2}
\usepackage{graphicx}
\usepackage{color}
\usepackage{braket}
\usepackage{amsmath}
\usepackage{amsfonts}
\usepackage{amsmath,amssymb,mathrsfs}
\usepackage{bbm}
\usepackage[dvipsnames]{xcolor}
\usepackage{appendix}
\usepackage{verbatim}
\usepackage{bbold}
\usepackage{footmisc}
\usepackage[normalem]{ulem}
\usepackage{xcolor}
\usepackage{soul}    
\usepackage[table]{xcolor}
\usepackage[unicode]{hyperref} 
\usepackage{blkarray}
\date{\today}

\begin{document}
	\title{\texorpdfstring{$\mathbb Z_{2q}$}{} parafermionic hinge states in a  three-dimensional array of coupled nanowires}

	\author{Sarthak Girdhar}
    \author{Viktoriia Pinchenkova}
    \author{Even Thingstad}
	\author{Jelena Klinovaja}
	\affiliation{Department of Physics, University of Basel, Klingelbergstrasse 82, CH-4056 Basel, Switzerland}
	\date{\today}

\begin{abstract}
We construct a model of a three-dimensional helical second-order topological superconductor formed by an array of weakly coupled Rashba nanowires. We identify the parameter regime in which there are energy gaps in both the bulk and surface energy spectra, while a pair of helical $\mathbb{Z}_{2q}$ parafermionic modes (with $q$ being an odd integer) remains gapless along a closed path of one-dimensional hinges. The precise trajectory of these hinge modes is dictated by the hierarchy of interwire couplings and the boundary termination of the sample. In the noninteracting limit $q= 1$, the system hosts gapless Majorana hinge modes.
\end{abstract}
\maketitle

\section{Introduction}\label{sec:Introduction}

\begin{figure}[t]
    \centering
\includegraphics[width=0.7\linewidth]{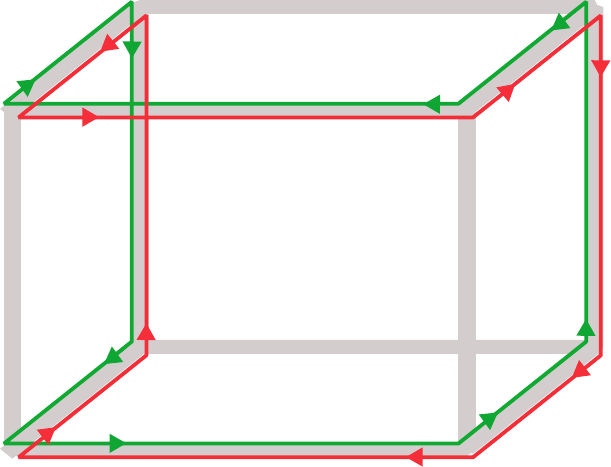}
    \caption{Schematic illustration of a helical second-order topological insulator/superconductor in three dimensions. The green and red lines represent Kramers pairs of gapless modes propagating along selected 1D hinges of the sample, depending on the parameters of the Hamiltonian. In contrast, the bulk and all surfaces of the system are fully gapped.}
 \label{fig:Schmatic_1}
\end{figure}

The classification of topological phases of matter has recently been expanded to include the so-called \textit{higher-order} topological phases~\cite{Benalcazar2017,Benalcazar2017b,Imhof2018,Song2017,Peng2017,Schindler2018,Geier2018,Langbehn2017}. 
Conventional topological insulators (TIs) and topological superconductors (TSCs) with a gapped $d$-dimensional bulk host gapless boundary modes of dimension $(d-1)$. In contrast, an $n$th-order TI or TSC supports protected gapless excitations only at their $(d-n)$-dimensional boundaries, while all higher-dimensional boundaries remain gapped. 
Among these, second-order topological phases have attracted particularly strong interest \cite{Second_Order_1,Second_Order_2,Second_Order_3,Second_Order_4,Second_Order_5,Second_Order_6}, as they give rise to zero-energy corner states in two-dimensional systems and gapless hinge states in three-dimensional ones [see, e.g., Fig.~\ref{fig:Schmatic_1}].

While most studies of higher-order TSCs (HOTSCs) and TIs have so far focused on noninteracting systems~\cite{Second_Order_5,Second_Order_6,Das_Sarma_Majorana_Hinge,Fu_Majorana_Hinge,Peng_Majorana_Hinge,Kheirkhah_Majorana_Hinge,Pahomi2019Braiding}, recent works have begun to explore how strong correlations may enrich this classification~\cite{You2018,You2019,Laubscher2019,Laubscher2020,May-Mann2022,Hackenbroich2021,Zhang2025_Construction,Li2022,Zhang2022b,Laubscher2023,Pinchenkova2025}.
The importance of this endeavor is underscored by the observation that many of the leading candidates for higher-order topological superconductivity are also characterized by properties typically associated with unconventional and strongly correlated superconductivity. 
A prime example is twisted bilayer graphene~\cite{Cao2018_Unconventional, Cao2018_CorrelatedInsulator}, where the flat bands ensure that electronic interactions dominate the physics~\cite{Bistritzer2011_Moire, Banerjee2025, Tanaka2025}. At the same time, the system has been proposed to host higher-order topological superconductivity~\cite{Chew2023_HigherOrder}. Moreover, it has also been proposed that WTe$_2$ hosts a higher-order topological superconducting phase with Majorana corner modes~\cite{Hsu2020_Inversion}. The violation of the Pauli (Chandrasekhar-Clogston) limit and the strong vortex Nernst signal in this system indicate that the pairing is unconventional~\cite{Sajadi2018_GateInduced, Asaba2018, Konov2021_WTe2, Song2024, Song2025_Unconventional}. 

One of the central challenges in understanding the role of interactions in HOTSCs is to describe strongly correlated phases analytically at the microscopic level, since electron-electron interactions must be treated in a nonperturbative manner. 
Among the few frameworks that enable the construction of analytically tractable toy models for strongly correlated topological phases is the \textit{coupled-wires} approach ~\cite{Kane2002,Kane2014}. 
In this method, higher-dimensional systems are built from arrays of weakly coupled one-dimensional (1D) nanowires, where intrawire electron-electron interactions can be incorporated naturally using standard bosonization techniques~\cite{Giamarchi}. 

Subsequently, interwire couplings are included perturbatively. 
This approach has proven remarkably powerful in the study of a wide range of exotic interacting first-order topological phases in two and three dimensions, including fractional quantum Hall states~\cite{Kane2002,Kane2014,Klinovaja2014b,Sagi2015b,Tam2021,Laubscher2021}, fractional quantum anomalous Hall states~\cite{Klinovaja2015,Liu2026}, fractional TIs~\cite{Klinovaja2014,Sagi2014,Neupert2014,Santos2015,Sagi2015, Meng2015}, chiral spin liquids~\cite{Gorohovsky2015, Meng2015_CSL, Thingstad2024}, and fractional TSCs~\cite{Neupert2014,Sagi2017,Li2020}. 
More recently, it has been shown that coupled-wires constructions can be exploited to study higher-order topological phases. 
However, only a few explicit examples of such models have been proposed so far~\cite{Laubscher2019,Laubscher2020,Zhang2022b,May-Mann2022,Laubscher2023,Pinchenkova2025}.

In this work, we introduce a three-dimensional (3D) coupled-wires model that realizes a rich variety of helical second-order topological superconducting (SOTSC) phases. In the noninteracting limit, the model hosts helical Majorana hinge states. While some previous works \cite{Second_Order_5,Second_Order_6,Das_Sarma_Majorana_Hinge,Fu_Majorana_Hinge,Peng_Majorana_Hinge,Kheirkhah_Majorana_Hinge,Pahomi2019Braiding} have also demonstrated such hinge states in noninteracting systems, our approach goes further by incorporating strong electron-electron interactions, leading to more exotic parafermionic hinge states. Unlike most known examples of HOTSCs, the stability of these states does not rely on specific spatial symmetries and is instead protected solely by particle-hole and time-reversal symmetry. Moreover, the parafermionic states we identify are topologically protected and exist at the hinges of a uniform 3D system, in contrast to previous works where they appear only at heterostructure interfaces \cite{Mong2014,Lindner2012,Clarke2013,Cheng2012,Orth2015,Fleckenstein2019} or at the ends of 1D nanowires \cite{Oreg_Fractional_Helical_Liquids,Klinovaja_Parafermion_Bundle,Thakurathi2017,Iemini2017,Calzona2018,Mazza2018}. Using perturbation theory, bosonization techniques, and numerical diagonalization, we further show how these HOTSCs can be systematically constructed from simpler and well-understood ingredients.

This paper is organized as follows. In Sec.~\ref{sec:Model}, we introduce the 3D coupled-wires model studied in this work. In Sec.~\ref{sec:Majorana}, we show that, for an appropriate choice of parameters, the model hosts gapless helical Majorana hinge states that propagate along a closed path around the 1D hinges of a finite 3D sample. In Sec.~\ref{sec:Parafermion}, we extend these results to the fractional case using bosonization techniques and demonstrate that sufficiently strong electron-electron interactions can give rise to a fractional helical SOTSC phase with gapless $\mathbb Z_{2q}$ parafermionic hinge states, where $q$ is an odd integer. Finally, we summarize our findings and provide an outlook in Sec.~\ref{sec:conclusion}.

\section{Model}\label{sec:Model}

\begin{figure}[t]
\centering
\includegraphics[width=0.96\linewidth]{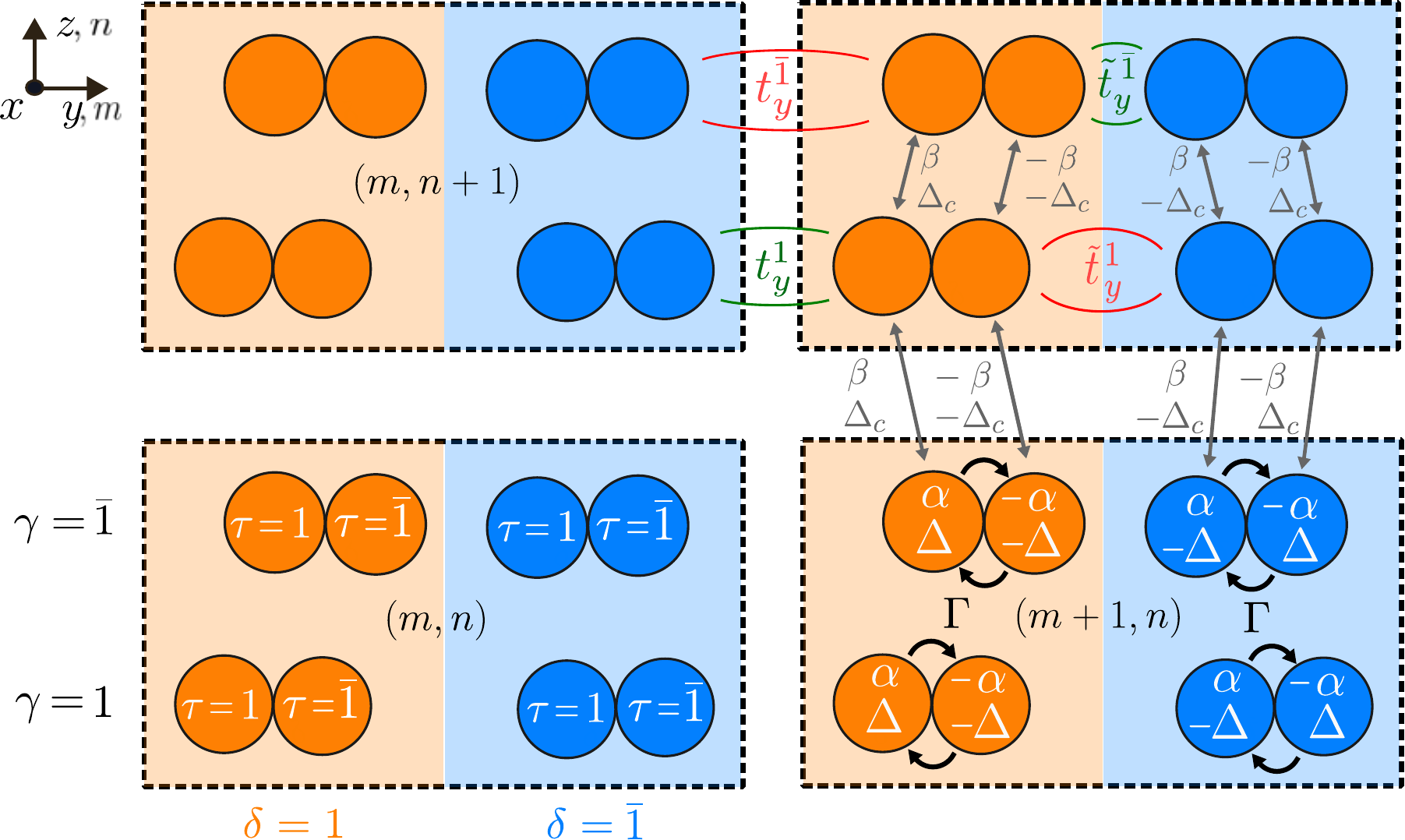}
\caption{Schematic of the 3D construction of 1D nanowires (circles) aligned along the $x$ axis. Each unit cell contains eight nanowires labeled $(m,n,\gamma,\delta,\tau)$, where $m$ and $n$ denote the cell position, and $\gamma,\delta,\tau\in\{1,\bar{1}\}$ distinguish between nanowires within each unit cell. Two nanowires with the same $\gamma$ and $\delta$ but different $\tau$ form a double nanowire (DNW), shown in orange ($\delta=1$) and blue ($\delta=\bar{1}$). Each wire is characterized by Rashba SOI of strength $\tau \alpha$ and proximity-induced superconductivity $\Delta$ with the phase $\delta \tau$. Along the $z$ direction, there is spin-flipping hopping of strength $\tau\beta$ and crossed-Andreev reflection $\delta \tau \Delta_c$. Along the $y$ axis, there is only spin-conserving hopping with amplitudes $\Gamma$ within a DNW and $t_y^\gamma$, $\tilde t_y^\gamma$ between DNWs, which are staggered such that $\Gamma$ (black) $\gg t_y^1, \tilde t_y^{\ \bar 1}$ (green) $\gg \tilde t_y^1, t_y^{\ \bar 1}$ (red). We assume that the values of these hopping amplitudes decrease with increasing distance between the nanowires.}
\label{fig:Schematic_Majorana_1} 
\end{figure}

In this section, we construct a model of a 3D SOTSC that can host Majorana and parafermionic hinge states. We start from
a 3D array of Rashba nanowires, shown in Fig.~\ref{fig:Schematic_Majorana_1}. Each circle represents a nanowire aligned along the $x$ direction. The nanowires are stacked in the $y$ and $z$ directions such that they form a unit cell consisting of eight nanowires. The position of a unit cell is labeled by the index $m$ along the $y$ direction and the index $n$ along the $z$ direction.

The nanowires within each unit cell are labeled by indices 
$\gamma, \delta, \tau \in \{1, \bar{1} \}$
[see Fig.~\ref{fig:Schematic_Majorana_1}]. 
We consider a model where the leading interwire coupling is within the pairs of nanowires indexed by $(\gamma,\delta)$ (adjacent circles),
and therefore refer to such a pair as a double nanowire (DNW).
As shown in Fig.~\ref{fig:Schematic_Majorana_1}, $\delta = 1$ ($\delta = \bar{1}$) therefore denotes the left (right) DNW within a given unit cell, 
$\gamma = 1$ ($\gamma = \bar{1}$) the bottom (top) DNW within a given unit cell, and $\tau = 1$ ($\tau = \bar{1}$) the left (right) wire within a given DNW. 

The total Hamiltonian of the system can be written as
\begin{equation}
H = H_0 + H_\Delta + H_\Gamma + \tilde{H}_z + H_z + \tilde{H}_y + H_y ,
\label{eq:Htotal}
\end{equation}
where $H_0$ contains the kinetic energy and strong spin--orbit interaction (SOI) terms, 
$H_\Delta$ describes the superconductivity in each nanowire, and 
$H_\Gamma$ accounts for tunneling between the two nanowires that form a DNW. 
The terms $H_z$ and $\tilde{H}_z$ correspond to the inter- and intra-unit-cell hopping 
processes along the $z$ direction, respectively, while $H_y$ and $\tilde{H}_y$ describe 
analogous tunneling processes along the $y$ direction. 
For the perturbative analysis introduced below, we assume a hierarchy of energy scales in 
which the chemical potential in each wire is the largest, followed by the intra-DNW 
tunneling, the tunneling along the $z$ direction, and finally the tunneling along the 
$y$ direction. In the following, we describe each of these terms in detail.

 For the sake of brevity, we introduce the composite DNW index $w=(m,n,\gamma,\delta)$. The first term describes the kinetic energy and the SOI through $H_0 = \sum_{w \tau\sigma} H_{0}^{w\tau\sigma}$, where the term describing spin $\sigma$ in the Rashba nanowire labeled by $(w,\tau)$ is given by 
\begin{equation} \label{eq:kinetic}
H_0^{w\tau\sigma}= \int dx\ \psi^{\dagger}_{w\tau\sigma}(x) \biggl(-\frac{\partial_x^2}{2m_e} -\mu+ i\tau\sigma \alpha \partial_x \biggr) \psi_{w\tau\sigma}(x),
\end{equation}
with $\psi^{\dagger}_{w\tau\sigma}(x)$ and $ \psi_{w\tau\sigma}(x)$ being the creation and annihilation operators for an electron at position $x$ in a wire $(w,\tau)$ with spin $\sigma \in \{1, \bar{1} \}$. The spin quantization axis is chosen along the $z$ direction. Here, $m_{e}$ is the effective electron mass, and we set $\hbar = 1$. 
Furthermore, $\alpha > 0$ is the Rashba spin-orbit strength, while the sign of the SOI depends on $\tau$, so that the two nanowires within a DNW have opposite spin structure [see Fig.~\ref{fig:ModeCouplingDNW}(a)].
The chemical potential is measured relative to the spin-orbit energy $E_{\text{so}} =  k_{\text{so}}^2 / 2 m_{e}$, where $k_\mathrm{so} = m_e \alpha$. Thus, as shown in Fig.~\ref{fig:ModeCouplingDNW}(a), for $\mu=0$, the Fermi momenta  are $k_F=0$ and $k_F=\pm 2 k_{so}$.

Next, we assume that the Rashba nanowires are subject to proximity-induced superconductivity with superconducting order parameter magnitude $\Delta > 0$ and sign alternating between the nanowires in a unit cell as shown in Fig.~\ref{fig:Schematic_Majorana_1}. The corresponding Hamiltonian is
\begin{equation}\label{proximityinducedsuperconductivity}
H_{\Delta} =\Delta \sum_{m,n}\sum_{\gamma,\delta,\tau} \delta \tau \int dx\ \psi_{w\tau 1}(x) \psi_{w\tau\bar{1}}(x) + \text{H.c.},
\end{equation} where the factor $\delta\tau$ encodes the aforementioned relative phase. Such a staggered phase difference can be achieved, for instance, by placing superconductors with alternating phases between DNWs in the $y$ direction, which in turn can be controlled by changing the magnetic flux through a superconducting loop~\cite{Josephson1,Fornieri2019} or by introducing a layer of randomly distributed scalar and randomly oriented spin impurities~\cite{PhysRevLett.115.237001} between the nanowires in a DNW.

We now start describing tunneling processes between the nanowires. We assume that the nanowires interact with their nearest neighbors in the $z$ direction through spin-flip hopping with amplitude $\beta$ and crossed-Andreev terms with amplitude $\Delta_c$. We further assume that these amplitudes are identical for both intracell and intercell interactions, and therefore, we have the following coupling terms:
\begin{align}
\tilde H_{z} &= \frac{1}{2}  \sum_{\substack{m,n,\\ \delta, \tau,\sigma, \sigma' }}  \int d x\ \Big( i \delta\tau \Delta_c
     \psi_{mn 1 \delta \tau \sigma}  \sigma_y^{\sigma \sigma'}   \psi_{mn\bar 1\delta\tau \sigma'}  \nonumber\\ &-  i\beta \tau \psi^{\dagger}_{mn 1 \delta \tau \sigma}    \sigma_x^{\sigma \sigma'}  \psi_{mn
     \bar 1\delta \tau \sigma'}+\ \text{H.c.} \Big), \label{intracellHz} \\
H_{z}&= \frac{1}{2}  \sum_{\substack{m,n,\\ \delta, \tau,\sigma, \sigma' }} 
\int dx\ \Big(
i \delta\tau \Delta_c
\psi_{mn \bar 1 \delta \tau \sigma} { \sigma_y^{\sigma \sigma'}}
\psi_{m(n + 1)1\delta\tau \sigma'} \nonumber \\ 
&- i \beta \tau \psi^{\dagger}_{mn \bar 1 \delta \tau \sigma} 
 \sigma_x^{\sigma \sigma'} 
\psi_{m(n + 1) 1\delta \tau \sigma'} 
+  \text{H.c.} 
\Big),
\label{intercellHz} \end{align} where $\sigma_j$ with $j \in \{x,y,z\}$ is a Pauli matrix for the spin degree of freedom. For brevity, we omit the explicit $x$-dependence of the field operators in the following. As given above, we assumed that the sign of the spin-flip hopping alternates with $\tau$ in the same way as the Rashba SOI [term $\propto \alpha$ in Eq.~(\ref{eq:kinetic})]. We further assumed that the sign of the crossed Andreev reflection term is given by $\delta \tau$, and coincides with the sign of the proximity-induced intrawire superconductivity in Eq.~(\ref{proximityinducedsuperconductivity}).

In the $y$ direction, we consider spin-conserving hopping described by five different hopping amplitudes as shown in Fig.~\ref{fig:Schematic_Majorana_1}. First, hopping between the two nanowires within a DNW is described by the amplitude $\Gamma$. 
Second, hopping between neighboring nanowires in different DNWs but within the same unit cell is characterized by the two $\gamma$-dependent amplitudes $\tilde{t}_{y}^{\gamma}$. 
Third, hopping between neighboring nanowires in different unit cells is described by the amplitudes $t_{y}^{\gamma}$. 
Altogether, hopping along the $y$ direction is therefore captured by the three contributions 
\begin{align}
\label{yintracellgammacoupling}
H_{\Gamma} &= \Gamma \sum_{\substack{m,n \\ \gamma, \delta,\sigma}} 
\int dx\ \psi^{\dagger}_{w 1 \sigma}\psi_{w\bar{1}\sigma} + \text{H.c.}, \\[2mm]
\label{yintracellnewcoupling}
\tilde{H}_{y} &= \sum_{\substack{m,n \\ \gamma,\sigma}} \tilde{t}_y^{\gamma} 
\int dx\ \psi^{\dagger}_{mn\gamma 1 \bar{1} \sigma}\psi_{mn\gamma \bar{1} 1 \sigma} + \text{H.c.}, \\[1mm]
\label{intercellnewcoupling}
H_{y} &= \sum_{\substack{m,n \\ \gamma,\sigma}} t_{y}^{\gamma} 
\int dx\ \psi^{\dagger}_{mn\gamma \bar{1} \bar{1} \sigma}\psi_{(m+1)n\gamma 1 1 \sigma} + \text{H.c.}
\end{align}
Equation~(\ref{eq:Htotal}) along with Eqs.~\eqref{eq:kinetic}--\eqref{intercellnewcoupling} constitutes the complete model. We emphasize that the model includes only nearest-neighbor tunneling processes in both the $y$ and $z$ directions. The system belongs to symmetry class DIII, characterized by the
presence of both time--reversal and particle--hole symmetries~\cite{ryu2010topological}. In addition, the bulk Hamiltonian respects several spatial symmetries, as listed in
Tab.~\ref{tab:spatial_symmetries}. However, as we show in
Sec.~\ref{subsec:numericalResults} and App.~\ref{appendix:disorder},
the hinge modes persist even when these spatial symmetries are broken, 
and their topological protection therefore does not rely on these symmetries.
\begin{table}
\caption{The list of spatial symmetries respected and broken by the bulk Hamiltonian.
Operations that reverse $z$ are broken by the $\gamma$-staggered interwire
hoppings ($t_y^1 \neq t_y^{\bar{1}}$, $\tilde{t}_y^1 \neq \tilde{t}_y^{\bar{1}}$). Protection of the hinge modes relies on time-reversal and particle-hole symmetries
(class DIII) and on none of the symmetries listed here.}
\label{tab:spatial_symmetries}

\begin{ruledtabular}
\begin{tabular}{llcc}
Operation & Type & Action on $(x,y,z)$ & Status \\
\hline
$\mathcal{M}_x$ & Mirror & $(-x,y,z)$ & preserved \\
$\mathcal{M}_y$ & Mirror & $(x,-y,z)$ & preserved \\
$C_{2z}$        & Rotation & $(-x,-y,z)$ & preserved \\
$\mathcal{M}_z$ & Mirror & $(x,y,-z)$ & broken \\
$\mathcal{I}$   & Inversion & $(-x,-y,-z)$ & broken \\
$C_{2x}$        & Rotation & $(x,-y,-z)$ & broken \\
$C_{2y}$        & Rotation & $(-x,y,-z)$ & broken \\
\end{tabular}
\end{ruledtabular}
\end{table}

In Secs.~\ref{sec:Majorana} and \ref{sec:Parafermion}, we explicitly demonstrate that the model introduced in this section can realize various 3D SOTSC phases characterized by a fully gapped bulk and fully gapped 2D surfaces, but hosting gapless helical hinge states that propagate along a closed path of 1D hinges.

\section{Helical Majorana Hinge Modes}\label{sec:Majorana}
In this section, we demonstrate that the Hamiltonian defined in Eq.~\eqref{eq:Htotal} can realize a SOTSC phase with helical Majorana hinge states in the noninteracting limit. The generalization to the interacting case with fractionalized parafermionic hinge states is discussed in Sec.~\ref{sec:Parafermion}.

To realize the helical Majorana hinge states, we tune the chemical potential $\mu$ to $0$, so that the two spin bands intersect at the Fermi level [see Fig.~\ref{fig:ModeCouplingDNW}(a)]. At this value of $\mu$, the interior branches of both
nanowires in a DNW cross at the same Fermi momentum $k_x=0$, such that the tunneling term $H_\Gamma$ conserves momentum exactly and can open a
gap without being dressed by electron-electron interactions
[see Sec.~\ref{sec:Parafermion}]. We emphasize that $\mu=0$ is not a
fine-tuned requirement: as we show in App.~\ref{appendix:disorder}, the hinge
modes persist for a finite range of chemical potentials around this point as long as the deviation of the chemical potential from $\mu=0$ stays smaller than the gaps opened in the spectrum.

We further assume that the system parameters obey the hierarchy
\begin{equation} \label{parameter_hierarchy} E_{\rm so} \gg \Delta, \Gamma \gg \beta, \Delta_c \gg t_{y}^{1}, 
\tilde{t}_{y}^{\,\bar{1}} \gg \tilde{t}_{y}^{1}, t_{y}^{\bar{1}} \ge 0, 
\end{equation}
which allows us to perform a multi-step perturbative procedure to include progressively smaller terms in the Hamiltonian.
 We first discuss uncoupled DNWs and identify their low-energy subspace. In Subsec.~\ref{subsec:DNWMajorana}, we show that each DNW hosts two Kramers pairs of gapless Majorana modes. In Subsec.~\ref{subsec:Majorana_hinge_analytics}, we include coupling between DNWs and demonstrate that these couplings lead to a fully gapped bulk energy spectrum and a fully gapped spectrum on the 2D surfaces, while two Kramers pairs of helical Majorana hinge states remain gapless. In Subsec.~\ref{subsec:yzSurface}, we consider nanowires of finite length $L$ in the $x$ direction and discuss hinge modes on the $yz$ surfaces. Then, in Subsec.~\ref{subsec:bulk_boundary_correspondence}, we recast these results in terms of a bulk-boundary correspondence, assigning a topological index to each gapped surface and showing that the hinge modes appear exactly where adjacent surface indices differ. Finally, in Subsec.~\ref{subsec:numericalResults}, we check the analytical results numerically by exact diagonalization and verify the presence of helical Majorana hinge modes. Moreover, we demonstrate that the strict parameter hierarchy we assumed to obtain the analytical results can be substantially relaxed as long as the bulk and the 2D surfaces remain gapped. 

\subsection{Majorana modes in uncoupled DNWs}\label{subsec:DNWMajorana}

For now, we assume that the system is infinite along the $x$ direction and consists of a unit cell repeated $N_y$  and $N_z$ times in the $y$ and $z$ directions, respectively. 
 
\begin{figure}[t]
    \centering  \includegraphics[width=0.8\linewidth]{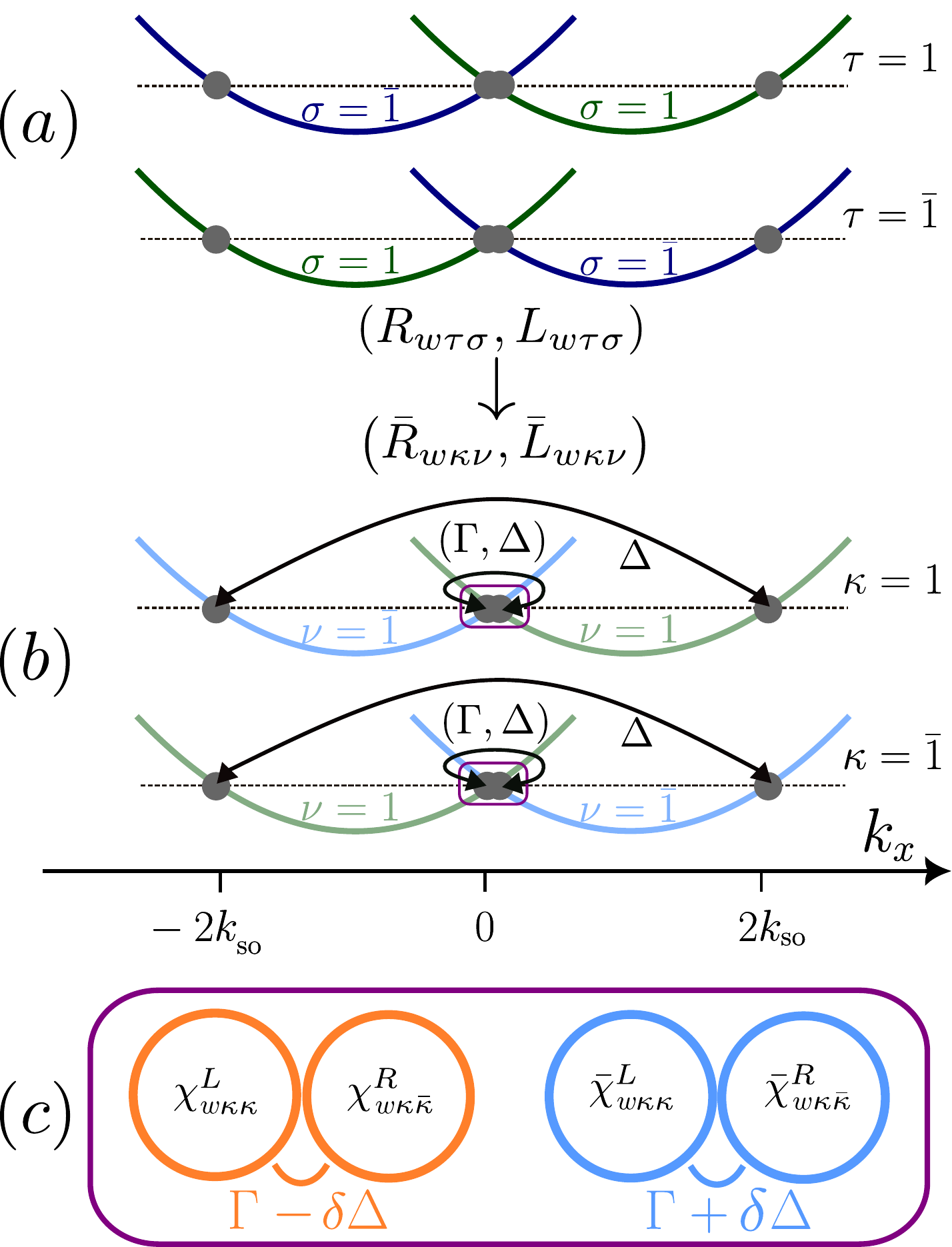}
    \caption{Spectrum of a DNW: (a) The DNW is composed of two Rashba nanowires (indexed by  $\tau \in \{1,\bar{1}\}$) with opposite spin structure (indexed by $\sigma \in \{1, \bar{1}\}$).
    (b) The transformation $\left(R_{w\tau\sigma},L_{w\tau\sigma}\right) \rightarrow \left(\bar R_{w\kappa\nu}, \bar L_{w\kappa\nu}\right)$  [cf. Eq.~\eqref{eq:unitary_transformation}], where  $w=(m,n,\gamma,\delta)$ is the DNW index, leaves the spectrum invariant. We can then label the bands in terms of the pseudolayer index $\kappa \in \{1,\bar 1\}$ and the pseudospin index $\nu \in \{1,\bar 1\}$. For each $\kappa$, the external modes at the Fermi momenta $\pm 2k_{\rm so}$ are gapped by the superconducting pairing $\Delta$. The internal modes (within purple boxes) are coupled by interlayer tunneling $\Gamma$ and proximity-induced superconductivity $\Delta$. (c) The two internal fermionic modes for a given $\kappa$ correspond to four Majorana modes $\chi_{w\kappa\kappa}^L$, $\chi_{w\kappa \bar{\kappa}}^R$, $\bar{\chi}_{w\kappa \kappa}^L$, $\bar{\chi}_{w\kappa\bar{\kappa}}^R$ [cf. Eq.~\eqref{majorana_R_L}]. For $\Gamma = \Delta$ and $\delta = 1$, the interplay between interlayer tunneling and superconductivity gaps out the modes $\bar{\chi}$ (blue outline), leaving the modes $\chi$ (orange outline) gapless. For $\delta = \bar{1}$, the situation is reversed: $\bar{\chi}$ are gapless and $\chi$ are gapped. 
}
\label{fig:ModeCouplingDNW}
\end{figure}

We first linearize the spectrum around the Fermi points by expressing the field operator $\psi_{w\tau\sigma}$ in terms of slowly varying right- and left-moving fields $R_{w\tau\sigma}$ and $L_{w\tau\sigma}$~\cite{Giamarchi}, so that
\begin{equation} \label{rightleftmover}
\psi_{w\tau\sigma}(x) = e^{i k_F^{1\tau\sigma} x} R_{w\tau\sigma}(x) + e^{i k_F^{\bar 1 \tau \sigma} x} L_{w\tau\sigma}(x),
\end{equation}
where the Fermi momentum is given by $k_F^{r\tau\sigma}=k_{\rm so}(r +\sigma\tau)$, with $r=1\,(r=\bar{1})$ denoting a right (left) mover. The possible values of the Fermi momentum are therefore zero and $\pm 2k_{\rm so}$ [see Fig.~\ref{fig:ModeCouplingDNW}(a)]. The right- and left-moving fields, $R_{w\tau\sigma}$ and $L_{w\tau\sigma}$, are assumed to vary slowly on the length scale of $k_{\rm so}^{-1}$.

First, we include only the intra-DNW coupling $\Gamma$ and the intra-wire proximity-induced superconducting gap $\Delta$, such that the full system decouples into noninteracting DNWs described by the Hamiltonian $H_0 + H_\Gamma + H_\Delta \equiv \sum_{(m,n,\gamma,\delta)} H_{\rm DNW}^{w}$. To determine the elementary excitations of $H_{\rm DNW}^{w}$, we perform the basis transformation
\begin{subequations}
\label{eq:unitary_transformation}
\begin{align} 
\label{unitary_transformation_R}
\bar{R}_{w \kappa \nu} &= \frac{1}{\sqrt{2}} \left( {R_{w\kappa \nu} - i \kappa \nu R_{w \bar{\kappa} \bar{\nu}}} \right), \\ 
\label{unitary_transformation_L}
\bar{L}_{w \kappa \nu} &= \frac{1}{\sqrt{2}} \left( {L_{w\kappa \nu} - i \kappa \nu L_{w \bar{\kappa} \bar{\nu}}} \right),
\end{align}
\end{subequations}
with pseudolayer index $\kappa$ and pseudospin index $\nu$, where  $\kappa, \nu \in \{1, \bar{1} \}$. 
Since the Fermi momentum $k_F^{r\tau\sigma}$ is invariant under $(\tau,\sigma) \rightarrow (\bar{\tau}, \bar{\sigma})$, the transformed modes can be associated with the unique Fermi momentum  $k_F^{r\kappa\nu} = k_{\rm so}(r+\kappa\nu)$. The DNW Hamiltonian then takes the form
\begin{align} \label{H_DNW_after_transformation}
H_{\rm DNW}^{w} &= -\frac{i\alpha}{2}\sum_{\kappa,\nu}\int dx \left( \bar R^\dagger_{w\kappa\nu}\partial_x \bar R_{w\kappa\nu} 
- \bar L^\dagger_{w\kappa\nu}\partial_x \bar L_{w\kappa\nu} \right) \nonumber \\
&\quad + \sum_{\kappa}  \int dx\, \left(i\Gamma \bar{R}^{\dagger}_{w\kappa \bar{\kappa}} \bar{L}_{w\kappa\kappa}  -  \delta  \Delta  \bar{R}^{\dagger}_{w\kappa\bar \kappa} \bar{L }^{\dagger}_{w\kappa\kappa} \right) \nonumber \\
&\quad + \sum_{\kappa} \delta  \Delta \int dx\, \bar{R}^{\dagger}_{w\kappa\kappa} \bar{L }^{\dagger}_{w\kappa\bar{\kappa}} + \text{H.c.}
\end{align}
This Hamiltonian is block-diagonal in $\kappa$, and we can therefore study the two blocks consisting of time-reversal partners separately. 

The exterior momentum branches $\bar R_{w\kappa\kappa}$ and $\bar L_{w\kappa\bar{\kappa}}$ at the respective Fermi momenta $k_F=+2k_{\rm so}$ and $k_F=-2k_{\rm so}$ are coupled only by the superconducting term given in the third line of Eq.~(\ref{H_DNW_after_transformation}), and therefore fully gapped [see Fig.~\ref{fig:ModeCouplingDNW}(b)]. In contrast, the interior branches $\bar R_{w\kappa\bar{\kappa}}$ and $\bar L_{w\kappa\kappa}$ are coupled both by the tunneling term $H_\Gamma$ and by the superconducting term $H_\Delta$ through the second line of Eq.~(\ref{H_DNW_after_transformation}). To analyze their competition, we express the interior branch operators in terms of Majorana operators $\chi^{R/L}_{w\kappa\nu}$ and $\bar{\chi}^{R/L}_{w\kappa\nu}$ through
\begin{subequations}
\label{majorana_R_L}
\begin{align}
\bar{R}_{w\kappa\bar{\kappa}} &= \frac{e^{i \pi /4}}{\sqrt{2}} 
\left( \chi^R_{w\kappa\bar{\kappa}} + i \bar{\chi}^R_{w\kappa\bar{\kappa}} \right), \\
\bar{L}_{w\kappa\kappa} &= \frac{e^{i \pi /4}}{\sqrt{2}} 
\left( \chi^L_{w\kappa\kappa} + i \bar{\chi}^L_{w\kappa\kappa} \right),
\end{align}
\end{subequations}
where expressions for $\left ( \bar{R}_{w\kappa\nu} \right)^\dagger$ and $\left ( \bar{L}_{w\kappa\nu} \right)^\dagger$ follow from the Majorana property $\chi^{R/L}_{w\kappa\nu} = \left(\chi^{R/L}_{w\kappa\nu}\right)^\dagger$ and $\bar{\chi}^{R/L}_{w\kappa\nu} = \left( \bar{\chi}^{R/L}_{w\kappa\nu}\right)^\dagger$.
Expressing the coupling between the interior modes in the second line of Eq.~\eqref{H_DNW_after_transformation} in terms of these operators, we find 
\begin{align}
    H_{\rm DNW,\, interior}^{w}
    &=i \sum_{\kappa}
    (\Gamma - \delta\Delta) \int dx\ \chi^R_{w\kappa \bar{\kappa}} \chi^L_{w \kappa
    \kappa} \nonumber \\ &+ i \sum_{\kappa} (\Gamma + \delta\Delta) \int dx\ \bar{\chi}^R_{w\kappa
    \bar{\kappa}} \bar{\chi}^L_{w \kappa \kappa}.
    \label{eq_hDNW_i}
\end{align}
Recalling the composite DNW index $w=(m,n,\gamma,\delta)$, we find that at the special point $\Gamma=\Delta$, there are two Kramers pairs of counterpropagating Majorana modes in each DNW. For DNWs with $\delta=1$, these are 
$\chi^R_{mn\gamma1\kappa\bar{\kappa}}$ and $ \chi^L_{mn\gamma1\kappa\kappa}$ with $\kappa \in \{1, \bar{1}\}$, 
while for $\delta=\bar 1$, the gapless modes are $\bar \chi^R_{mn\gamma \bar 1\kappa\bar{\kappa}}$ and $\bar \chi^L_{mn\gamma \bar 1\kappa\kappa}$ [see Fig.~\ref{fig:ModeCouplingDNW}(c)]. For convenience, we therefore introduce the notation
\begin{subequations}
\label{eq_chiRedefM}
  \begin{align}
\chi^{\kappa R}_{mn \gamma\delta} =\begin{cases}\chi^R_{mn\gamma 1\kappa\bar\kappa} & \text{for }\delta=1 \\ \bar \chi^R_{mn\gamma \bar 1 \kappa\bar\kappa}  & \text{for } \delta=\bar 1\end{cases},
\\
\chi^{\kappa L}_{mn \gamma\delta} =\begin{cases}\chi^L_{mn\gamma 1\kappa\kappa} & \text{for }\delta=1 \\ \bar \chi^L_{mn\gamma \bar 1 \kappa \kappa}  & \text{for } \delta=\bar 1\end{cases}.
\end{align}
\end{subequations}
To recapitulate, we started out with eight fermionic modes in a DNW [see Fig.~\ref{fig:ModeCouplingDNW}(a)], which correspond to $16$ Majorana modes. The eight Majorana modes corresponding to the exterior branches of the dispersion relation are gapped out by the superconducting term [see Fig.~\ref{fig:ModeCouplingDNW}(c)]. Of the remaining eight modes, half of them are gapped due to the couplings in Eq.~\eqref{eq_hDNW_i}. For a given DNW indexed by $(m,n,\gamma,\delta)$, this leaves two pairs of Majorana modes $\chi_{mn\gamma\delta}^{\kappa R}$ and $\chi_{mn\gamma\delta}^{\kappa L}$ gapless [see Fig.~\ref{fig:ModeCouplingDNW}(b)]. 

At the next step, we want to gap out these initially gapless Majorana modes by coupling each DNW to the neighboring DNWs. We will demonstrate that the Majorana modes on DNWs located in the bulk or on the 2D surfaces are gapped out, while the Majorana modes located on the hinges stay gapless. Due to the presence of time-reversal symmetry, these Majorana modes are helical.
  
\subsection{Majorana hinge modes in the 3D SOTSC} \label{subsec:Majorana_hinge_analytics}

In the previous Subsection, we showed that all DNWs host two pairs of counterpropagating gapless Majorana modes each. We now switch on these couplings  $\beta,\Delta_c \gg t_{y}^1,\tilde t_{y}^{\, \bar{1}} \gg  t_{y}^{\bar 1},\tilde t_{y}^{\, 1}$,  and examine their effect on these gapless modes.

\begin{figure}[ht!]
    \centering
\includegraphics[width=0.95\linewidth]{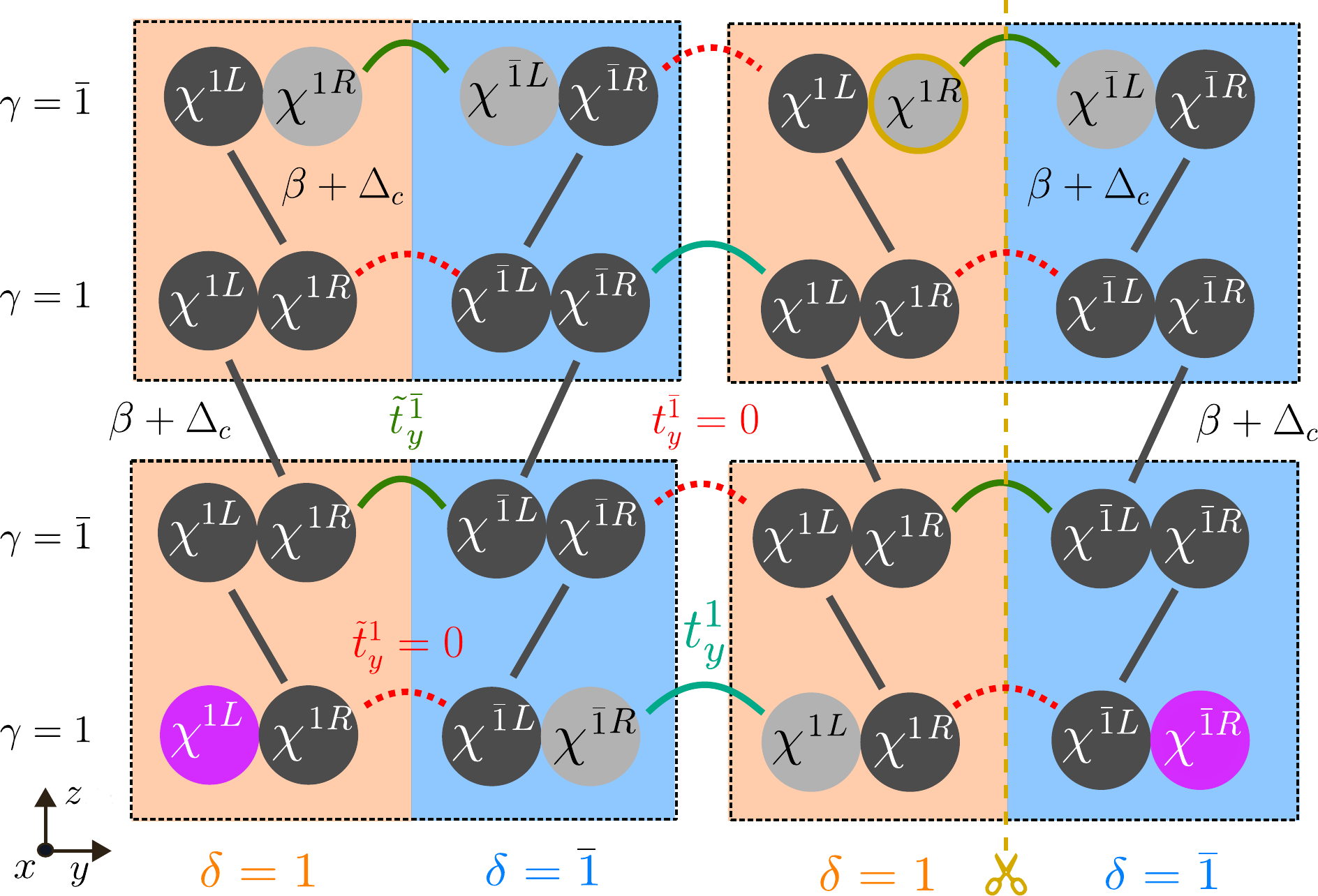}
    \caption{Coupling of Majorana modes for the time-reversal sector corresponding to modes with $\kappa=1$ for $\delta=1$ and $\kappa=\bar{1}$ for $\delta= \bar{1}$ due to inter-DNW coupling for a system of $N_y \times N_z=2\times 2$ unit cells. As shown in Fig.~\ref{fig:ModeCouplingDNW}, each DNW hosts two Kramers pairs of gapless Majorana modes $\chi^{\kappa R}, \chi^{\kappa L}$, where $\kappa\in \{1,\bar{1}\}$ is the pseudolayer index. Out of these four modes, we show the two modes corresponding to a given time-reversal sector as a pair of circles for every DNW. Crossed Andreev reflection and spin-flip hopping in the $z$ direction (solid dark grey lines) gap out the bulk modes (dark grey circles) and the $xz$ surfaces. Hopping in the $y$ direction further gaps out the remaining $xy$ surface modes (light grey circles), leaving gapless Majorana modes $\chi^{1L}_{1111}$ and  $\chi^{\bar 1R}_{N_y11 \bar 1}$ (magenta circles) at the hinges of the system.
    Changing the boundary termination by removing the rightmost DNW layer (indicated by scissors) relocates the gapless hinge state from $\chi^{\bar 1R}_{N_y11 \bar 1}$ to $\chi^{1R}_{N_yN_z \bar 11}$ (circle with yellow outline). The time-reversed partner of the above figure can be found by replacing $\chi^{1R(L)} \leftrightarrow \chi^{\bar 1 L(R)}$.
    } \label{fig:ModeCouplingMajoranaChi}
\end{figure}

We start with the terms in the Hamiltonian $H$ [see Eq.~\eqref{eq:Htotal}] that couple neighboring DNWs in the $z$ direction. We consider only the terms affecting the low-energy modes 
$\{\chi^{\kappa R}_{mn\gamma 1},\ \chi^{\kappa R}_{mn\gamma \bar{1}},\ \chi^{\kappa L}_{mn\gamma 1},\ \chi^{\kappa L}_{mn\gamma \bar{1}}\}$.
This gives
\begin{subequations} \label{z_terms_majorana}
\begin{align}
\tilde H_{z} &= \frac{i}{2} \sum_{m,n,\delta,\kappa} \int dx \, \Big[
      (\kappa\beta - \Delta_c) 
\chi^{\kappa R}_{mn \bar 1 \delta} \chi^{\kappa L}_{mn1 \delta}
      \nonumber \\
 & +  (\kappa\beta + \Delta_c) 
\chi^{\kappa L}_{mn \bar 1 \delta} \chi^{\kappa R}_{mn1 \delta}
 \Big], \label{intra_Hz_as_Majorana} \\
H_{z} &= \frac{i}{2} \sum_{m,\delta,\kappa} \sum_{n=1}^{N_z-1} \int dx \, \Big[
    (\kappa\beta - \Delta_c) 
\chi^{\kappa R}_{m(n+1)1\delta} \chi^{\kappa L}_{mn\bar 1\delta} \label{inter_Hz_as_Majorana}
    \nonumber \\
&+ (\kappa\beta + \Delta_c) 
\chi^{\kappa L}_{m(n+1)1\delta} \chi^{\kappa R}_{mn\bar 1\delta}
\Big].
\end{align}
\end{subequations}
To simplify the discussion, we consider the special point $\beta=\Delta_c$ 
in the following. Yet, in App.~\ref{appendix:disorder}, we check numerically that the qualitative results we obtain below hold even if this is not the case. With the given assumption, the gapless modes
$\{\chi^{1L}_{m111}, \chi^{1R}_{mN_z\bar 1 1}, \chi^{1L}_{m11 \bar 1}, \chi^{1R}_{mN_z\bar 1 \bar 1}\}$
and their time-reversed partners 
$\{\chi^{\bar 1R}_{m11 1},\ \chi^{\bar 1L}_{mN_z\bar 1 1}, \chi^{\bar 1R}_{m11 \bar 1},\ \chi^{\bar 1L}_{mN_z\bar 1 \bar1}\}$
do not enter the Hamiltonian, and thus remain gapless. These gapless states are located on the $xy$ surfaces of the sample. On the other hand, all other states in the bulk and on
the $xz$ surfaces of the sample are fully gapped out (cf. black circles in Fig.~\ref{fig:ModeCouplingMajoranaChi}).

In the last step of the perturbation procedure, we introduce coupling in the $y$ direction to gap out all modes on the $xy$ surfaces except for the hinge modes. Expressing the hopping in the $y$ direction in terms of the Majorana modes which so far remained gapless, we have
\begin{subequations}\label{eq:hy}
\begin{align} \label{finalformintracell}
\tilde H_{y} &= \frac{i}{2} 
    \sum_{n\in \{1, N_z\}}
    \sum_{m,\gamma,\kappa} \kappa \, \tilde t_y^{\gamma} \int dx \,   \Big[
   \chi^{\bar{\kappa} R}_{mn\gamma\bar{1}}\chi^{\kappa L}_{mn\gamma 1}
    \\ \nonumber 
&\quad + 
\chi^{\bar\kappa L}_{mn\gamma\bar1}\chi^{\kappa R}_{mn\gamma1}
\Big], \\
\label{ycouplingfinalintercell}
H_y &= \frac{i}{2} 
    \sum_{n\in \{1, N_z\}}
    \sum_{m=1}^{N_y-1} 
    \sum_{\gamma,\kappa} \kappa \,
    t_y^{\gamma} \int dx \, \Big[
    \chi^{\kappa R}_{(m+1)n\gamma 1}  {\chi}^{\bar \kappa L}_{mn\gamma\bar 1} 
    \nonumber \\
&\quad + 
\chi^{\kappa L}_{(m+1)n\gamma 1}\chi^{\bar \kappa R}_{mn\gamma\bar 1} 
\Big].
\end{align}
\end{subequations}
The above interaction couples almost all the modes on the two $xy$ surfaces of the sample (cf. green couplings in Fig.~\ref{fig:ModeCouplingMajoranaChi}). Yet, as shown in magenta, there are gapless states localized on the two hinges $(m,n) = (1,1)$ and $(m,n)=(N_y,1)$: the left- and right-movers    $(\chi^{1L}_{1111}, \chi^{\bar 1R}_{N_y11 \bar 1})$ and their time-reversed partners $(\chi^{\bar 1R}_{111 1}, \chi^{ 1L}_{N_y11 \bar 1})$. Hence, we have shown that for infinite nanowires, in a suitable parameter regime, our model hosts Majorana modes at the hinges of the 3D sample. It should be noted that the strict conditions $\Gamma=\Delta$ and $\beta=\Delta_c$ can be relaxed, provided that the deviations $|\Gamma-\Delta|$ and $|\beta-\Delta_c|$ remain smaller than the corresponding subsequent energy scales in our parameter hierarchy. In that case, these smaller scales are still sufficient to gap out the required low energy Majorana modes in the bulk and on the surfaces of the 3D sample.

\subsection{Hinge modes on the \texorpdfstring{$yz$}{} surfaces}\label{subsec:yzSurface}
 So far, we have shown that for infinite nanowires, the system can realize a SOTSC phase with a fully gapped bulk, fully gapped $xy$ and $xz$ surfaces, and two Kramers pairs of gapless Majorana hinge modes propagating along the $x$ direction. In the following, we address whether the $yz$ surfaces appearing at $x=0$ and $x=L$, where $L$ is the length of finite nanowires, also host hinge modes. 

In the absence of inter-DNW hopping amplitudes $\left(t_y^1,\, t_y^{\bar{1}},\, \tilde{t}_y^{\, 1},\, \tilde{t}_y^{\, \bar{1}}\right)$ along the $y$ direction, the 3D model reduces to decoupled two-dimensional bilayers stacked along the $y$ axis. In the parameter regime $E_{\rm so} \gg \Gamma,\, \Delta \gg \beta,\, \Delta_c$, it has been shown that such an independent bilayer realizes a topological superconducting phase with helical Majorana modes at the edges~\cite{Laubscher2019}. 

\begin{figure*}[]
    \centering
    \includegraphics[width=\textwidth]{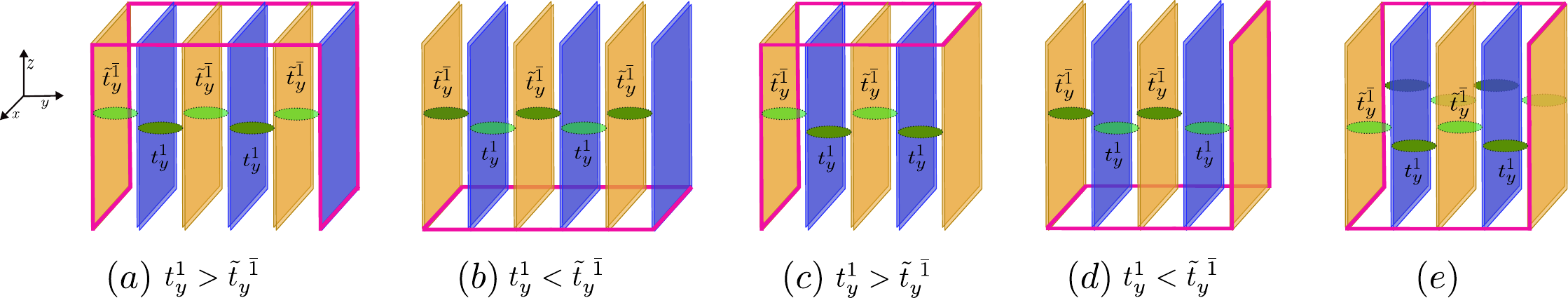}
\caption{
Model viewed as a stack of bilayers (orange: $\delta=1$, blue: $\delta=\bar 1$) coupled along the $y$ axis.
Each uncoupled bilayer hosts Majorana edge modes. Turning on the tunnelings $\tilde t_y^{\,\bar 1}$ and $t_y^{1}$ gaps out the surface modes, while hinge modes may remain gapless. The dimerization pattern of the bilayers is set by which of $\tilde t_y^{\,\bar 1}$ or $t_y^{1}$ is larger (dark green) or smaller (light green). In the sector $\kappa=1$, for the  patterns in (a) and (b), the hinge modes $\chi^{1L}_{1111}$ and $\chi^{\bar 1 R}_{N_y 11\bar 1}$ propagate along the $x$ direction, and are connected via the $y$ hinges  of the upper and lower surface, respectively. In (c) and (d), removing the rightmost bilayer, which breaks the mirror symmetry $\mathcal M_y$ (see Tab.~\ref{tab:spatial_symmetries}),replaces $\chi^{\bar 1 R}_{N_y 11\bar 1}$ with $\chi^{1R}_{N_y N_z \bar 1 1}$, yielding a different hinge geometry. In (e), opposite dimerization patterns for $x<L/2$ and $x>L/2$, which further break the mirror symmetry $\mathcal M_x$ (see Tab.~\ref{tab:spatial_symmetries}), produce the indicated hinge mode configuration. As stated earlier, the modes persist even after these spatial symmetries are broken, and their robustness relies solely on particle-hole and time-reversal symmetry. We verify the above hinge mode geometries numerically in Fig.~\ref{fig:density}.
\label{fig:layer_stack}}
\end{figure*}

The interactions in Eq.~\eqref{eq:hy} couple these modes directly. To examine the effect of the competing couplings $\tilde{t}_y^{\, \bar{1}}$ and $t_y^1$ on gap formation at the $yz$ surfaces, we evaluate the overlap integrals between low energy edge states of adjacent bilayers. 
We impose periodic boundary conditions in the $z$ direction and label the states propagating in this direction by the quasimomentum $k_z$, so that the quantum state of the bilayer with index $\delta$ in the $m^{\mathrm{th}}$ unit cell is denoted by 
$|\bar{\Psi}^{\kappa k_z}_{m\delta} \rangle$~\cite{Note2}, where the states are doubly degenerate ($\kappa \in \{1, \bar{1}\}$) due to the time-reversal symmetry of the model. 
For a single bilayer, we find the zero energy modes at quasimomentum $k_z=\pi/a_z$, where $a_z$ is the nearest neighbor wire distance in the $z$ direction. Explicit expressions for the eigenstates are given in App.~\ref{appendix:finite_nanowire_states}. 
To find out whether hopping in the $y$ direction opens gaps on the $yz$ surfaces, we calculate the matrix elements
\begin{subequations}
\begin{align}
\label{intracell_y_overlap}
 \left\| \left\langle \bar{\Psi}^{\kappa \pi}_{m1} \Big| \tilde{H}_y \Big| \bar{\Psi}^{\bar\kappa \pi}_{m \bar{1}}\right\rangle  \right\| & = c\, \tilde{t}_y^{\, \bar{1}},
\\
\label{intercell_y_overlap}
    \left\| \left\langle 
    \bar{\Psi}^{\bar{\kappa} \pi}_{m\bar{1}} \Big| H_y \Big| 
    \bar{\Psi}^{\kappa \pi}_{(m+1) 1 }
    \right\rangle \right\| 
    & = c\, t_y^{1},
\end{align}
\end{subequations}
where $c$ is a common positive constant, as shown in App.~\ref{appendix:finite_nanowire_states}. 
Since $k_z$ is a good quantum number when the system is infinite (or periodic) in the $z$ direction, only modes at the same $k_z$ can couple. Thus, the low-energy Hamiltonian decouples into two ($\kappa \in \{ 1, \bar{1}\}$) effective SSH models composed of the states $\{\bar{\Psi}_{11}^{\kappa\pi}, \bar{\Psi}_{1\bar{1}}^{\bar{\kappa}\pi}, \bar{\Psi}_{21}^{\kappa\pi}, \dots, \bar{\Psi}_{N_y \bar{1}}^{\bar{\kappa}\pi}\}$, where the index $\delta$ labels the two inequivalent sites of the effective SSH model.
Based on this insight, we conclude that the modes on the $yz$ surface remain gapless when the tunneling amplitudes satisfy $t_{y}^1 = \tilde{t}_{y}^{\, \bar 1}$. When the tunneling amplitudes differ, the $yz$ surfaces become fully gapped, and only the hinge states may remain gapless. The precise location of these hinge states depends on the dimerization pattern responsible for opening the surface gaps [see Fig.~\ref{fig:layer_stack}], as determined from the effective SSH model. For dominant intercell coupling ($t_y^1 > \tilde{t}_{y}^{\, \bar{1}}$), the $yz$ surfaces are gapped out like in the topologically nontrivial SSH model, yielding four hinge modes propagating along the $z$ direction which can only be connected on the upper surface through two hinge modes propagating along the $y$ direction [see Fig.~\ref{fig:layer_stack}(a)]. Conversely, for dominant intracell coupling, ($t_y^1 < \tilde{t}_{y}^{\, \bar{1}}$), the $yz$ surfaces are gapped out like in the topologically trivial SSH model, and no hinge states propagate along the $z$ direction. The resulting hinge mode geometry is indicated in Fig.~\ref{fig:layer_stack}(b).

While the above arguments rely on a multi-step perturbation theory, which is strictly speaking only valid when the system parameters satisfy the given hierarchy, our  qualitative results remain valid as long as the bulk and surface gaps do not close through tunnelings due to the Hamiltonians $H_y$ and $\tilde H_y$. Thus the  requirement $\beta, \Delta_c \gg t_{y}^1, \tilde{t}_{y}^{\, \bar{1}}$ can be relaxed. This can be checked numerically, as we show in Sec.~\ref{subsec:numericalResults}.
\begin{table}[t]
\caption{System parameters used to obtain the tight-binding results in Fig.~\ref{fig:density}. The details of the discretization can be found in App.~\ref{appendix:tight_binding_details}. The hopping matrix element in the $x$ direction is $t_x=1/(2m_ea_x^2)=0.9E_{\rm so}$, while $\alpha/2a_x=0.95E_{\rm so}$, where $a_x$ is the lattice constant in the $x$-direction. We further consider $t_y^{\bar{1}} = \tilde{t}_y^1 = 0$. The system has $N_y \times N_z=15\times10$ unit cells and $N_x = 80$ sites in the $x$ direction. Density plots in the presence of random onsite charge disorder at $\mu=0$, including deviations from the fine-tuned conditions $\Gamma = \Delta$ and $\beta = \Delta_c$ (see below), are shown in App.~\ref{appendix:disorder}.}

\begin{ruledtabular}\label{tab:parameters}
\begin{tabular}{ccccccc}
Plot  & $\Gamma$ & $\Delta$ & $\beta$ & $\Delta_c$ & $t_y^1$ & $\tilde{t}_y^{\, \bar{1}}$ \\
\hline
(a)                & 0.60 & 0.60 & 0.25 & 0.25 & 0.40 & 0.18 \\
(b)            & 0.60 & 0.60 & 0.25 & 0.25 & 0.18 & 0.40 \\
(c)               & 0.60 & 0.60 & 0.25 & 0.25 & 0.40 & 0.18 \\
(d)              & 0.60 & 0.60 & 0.25 & 0.25 & 0.18 & 0.40 \\
(e), $x<L/2$     & 0.60 & 0.60 & 0.25 & 0.25 & 0.18 & 0.40 \\
(e), $x>L/2$       & 0.60 & 0.60 & 0.25 & 0.25 &  0.40 & 0.18 \\
\end{tabular}
\end{ruledtabular}
\end{table}

\begin{figure*}[]
    \centering
    \includegraphics[width=01\textwidth]{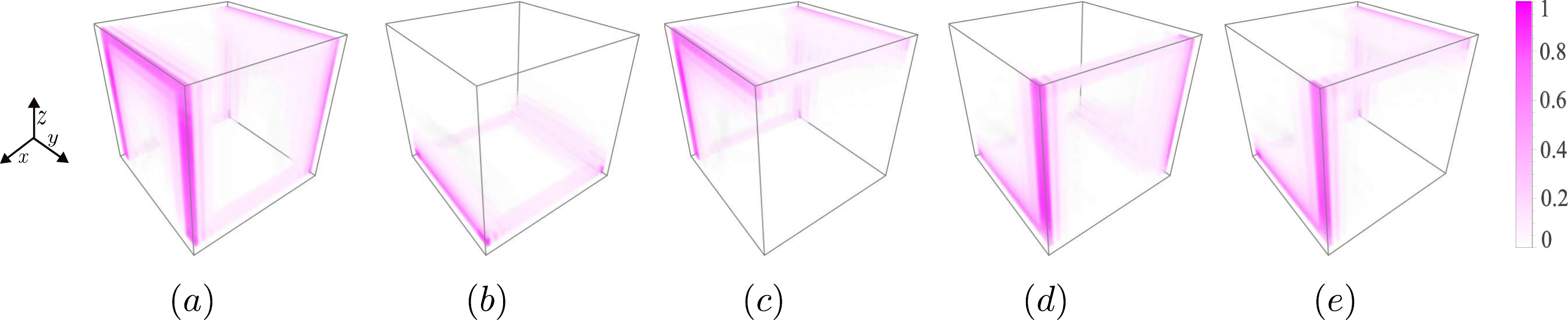}
    \caption{Probability density for the lowest-energy eigenstate, obtained by exact diagonalization of a discretized version of Eq.~(\ref{eq:Htotal}) with parameter values as in Table~\ref{tab:parameters}. In all cases, Majorana hinge states (magenta) form closed paths consistent with the dimerization patterns and boundary terminations in Fig.~\ref{fig:layer_stack}. The color bar represents the probability at each site, normalized such that the most probable site has value 1. The system size is $15 \times 10$ unit cells ($60 \times 20$ sites), with a wire having $N_x = 80$ sites.} \label{fig:density}
\end{figure*}

\subsection{Bulk-boundary correspondence and surface topological invariants}
\label{subsec:bulk_boundary_correspondence}
Having established the presence of hinge modes analytically, we now show that the same geometries can also be viewed from the perspective of a bulk-boundary correspondence. This line of argument both explains why the helical modes are localized on the particular hinges shown in
Fig.~\ref{fig:layer_stack} and establishes their connection to surface topological invariants. The essential
point is that all 3D bulk invariants of our phases are identically zero, so their nontrivial topology must instead be encoded in the gapped two-dimensional surfaces~\cite{Kane2005first,FuKaneMele2007, QiHughesRaghuZhang2009}.

We first confirm that all first-order invariants vanish. From Sec.~\ref{subsec:yzSurface} and Eq.~\eqref{parameter_hierarchy}, we observe that the 3D bulk Hamiltonian of our system is adiabatically connected to that of a
stack of decoupled two-dimensional bilayers along the $y$ direction, which is independent of the quasimomentum $k_y$. This implies that the strong class-DIII bulk winding number~\cite{PhysRevB.78.195125,MooreBalents2007,ryu2010topological,KaneInversionSymmetry,Roy2009} vanishes. Similarly, the weak bulk invariant also vanishes: each unit cell contains two
bilayers ($\delta = 1, \bar{1}$), each of which is a two-dimensional topologically nontrivial class-DIII superconductor~\cite{Laubscher2019} with a $\mathbb Z_2$ index~\cite{Qi2010} equal to $-1$. Under the hierarchy of Eq.~\eqref{parameter_hierarchy}, the couplings $t_y^\gamma$ and $\tilde t_y^\gamma$ cannot change the 2D bulk properties of these two bilayers. Since both bilayers are nontrivial, their combined contribution to the weak topological invariant is trivial. The phase is therefore invisible to every
first-order invariant and is genuinely of second order, its topology residing in
the $\mathbb{Z}_2$ indices of the gapped surfaces.

To assign such an index to a given surface, we observe from Fig.~\ref{fig:ModeCouplingMajoranaChi} and Sec.~\ref{subsec:yzSurface} that once the bulk and the
remaining surface modes have been gapped, the $xy$ and $yz$ surfaces of our 3D system can be viewed as effective SSH models. This means that the couplings acting within one of these surfaces gap it in a dimerized pattern, and the competition between its intra- and
inter-cell amplitudes fixes a class-DIII surface index $\nu_F = \pm 1$, with $+1$
($-1$) denoting the trivial (topological) phase. The surface is thus topological when its inter-cell coupling dominates. As
discussed in Sec.~\ref{subsec:yzSurface}, the $yz$ surfaces are governed by
the amplitudes $\tilde{t}_y^{\bar{1}}$ and $t_y^1$ and are topological when the
inter-cell coupling $t_y^1$ dominates. The two $xy$ surfaces are likewise described by effective SSH
chains, but they are not equivalent: 
the bottom $xy$ surface exposes the $\gamma=1$ sublattice, while the top surface exposes the $\gamma = \bar{1}$ sublattice. 
 
Because the hoppings $t_y^\gamma$ and $\tilde t_y^\gamma$ are staggered with respect to $\gamma$, these two $xy$ surfaces inherit opposite
dimerization patterns, so that under Eq.~\eqref{parameter_hierarchy} the bottom surface is
topological while the top surface is trivial. Finally, both $xz$ surfaces are topologically trivial for the same reason that the weak bulk invariant vanishes, as discussed earlier in this subsection. However, a simple way to make the left (right) $xz$ surface topological is by removing a bilayer from the first (last) unit cell in Fig.~\ref{fig:Schmatic_1}. We show this in Figs.~\ref{fig:layer_stack}(c)--(e), and list the resulting indices for all configurations of
Fig.~\ref{fig:layer_stack} in Tab.~\ref{tab:invariants}.

\begin{table}
\caption{Second-order class-DIII $\mathbb{Z}_2$ surface invariants
$\nu_F = \pm 1$ for the configurations of Fig.~\ref{fig:layer_stack}. All
first-order bulk invariants vanish, and a gapless hinge mode appears only
between surfaces with unequal invariants.}
\label{tab:invariants}

\begin{ruledtabular}
\begin{tabular}{ccccccc}
 Surface $\rightarrow$ & $y=0$ & $y=N_y$ & $x=0$ & $x=N_x$ & $z=0$ & $z=N_z$ \\
\hline
5(a) & 1 & 1 & -1 & -1 & -1 & 1 \\
5(b) & 1 & 1 & 1 & 1 & -1 & 1 \\
5(c) & 1 & -1 & -1 & -1 & -1 & 1 \\
5(d) & 1 & -1 & 1 & 1 & -1 & 1 \\
5(e) & 1 & -1 & 1 & -1 & -1 & 1 \\
\end{tabular}
\end{ruledtabular}

\end{table} 

The hinge spectrum now follows from a single rule. In the spirit of the
Jackiw-Rebbi/Teo-Kane domain-wall mechanism~\cite{Jackiw1976,TeoKane2010}, a
hinge shared by surfaces $F$ and $F'$ is a 1D domain wall between two
gapped surface superconductors and binds a gapless helical hinge mode if and
only if the two surface indices differ, i.e. $\nu_F\,\nu_{F'} = -1$. Together with
Table~\ref{tab:invariants}, this criterion reproduces every hinge configuration
of Fig.~\ref{fig:layer_stack}.

\subsection{Numerical results}\label{subsec:numericalResults}
We verify our analytical predictions using exact diagonalization in the tight-binding limit. The computational procedure is described in detail in App. \ref{appendix:tight_binding_details}, and the resulting low energy mode probability density profiles are shown in Fig.~\ref{fig:density} for the parameters listed in Table~\ref{tab:parameters}. In Fig.~\ref{fig:density}(a), we show the probability density for a Majorana fermion in the lowest energy state for a set of parameters with $t_y^1 > \tilde{t}_{y}^{\, \bar{1}}$, so that we expect four hinge modes propagating along the $z$ direction, as discussed above. This is exactly what we find. As expected, in Fig.~\ref{fig:density}(b), we find no hinge modes propagating along the $z$ direction when $t_y^1 < \tilde{t}_{y}^{\, \bar{1}}$. 
The hinge-state configuration is also sensitive to the boundary termination, as one would expect based on the SSH model. For instance, by removing the rightmost bilayer lying in the $xz$ plane at $m = N_y$, $\delta = \bar{1}$, the hinge modes propagating in the $z$ direction vanish on the right end of the system, 
while the hinge state propagating along $x$ relocates from the DNW $(m,n,\gamma,\delta) = (N_y, 1,1, \bar{1})$ to $(N_y, N_z, \bar 1,1)$ [see Fig.~\ref{fig:ModeCouplingMajoranaChi}]. The hinge states along other directions adapt accordingly, forming a closed path consistent with the underlying dimerization pattern. As a result, we obtain the hinge mode geometries in Figs.~\ref{fig:density}(c) and~\ref{fig:density}(d) in the respective regimes $t_y^1 > \tilde{t}_y^{\,\bar{1}}$ and $t_y^1 < \tilde{t}_y^{\,\bar{1}}$. In Fig.~\ref{fig:density}(e), we consider a geometry where the hopping amplitudes $t_y^1$ and $\tilde t_y^{\,\bar 1}$ are taken to depend on $x$ in such a way that the system has one dimerization pattern for $x< L/2$, and another for $x>L/2$. Again, we find that the numerical results agree with what we expect from the SSH model. Although this parameter choice is rather artificial, it serves to illustrate the considerable flexibility of our model: by appropriately tuning the system parameters, a diverse set of hinge states can be achieved. In App. \ref{appendix:disorder}, we show that the hinge states persist even when the system is detuned from the fine-tuned conditions $\Gamma=\Delta$,  $\beta=\Delta_c$, $\mu=0$, and when onsite charge disorder is introduced. As expected, they exhibit excellent stability under both types of deviations.

\begin{figure}[htb]
    \centering
 \includegraphics[width=0.95\linewidth]{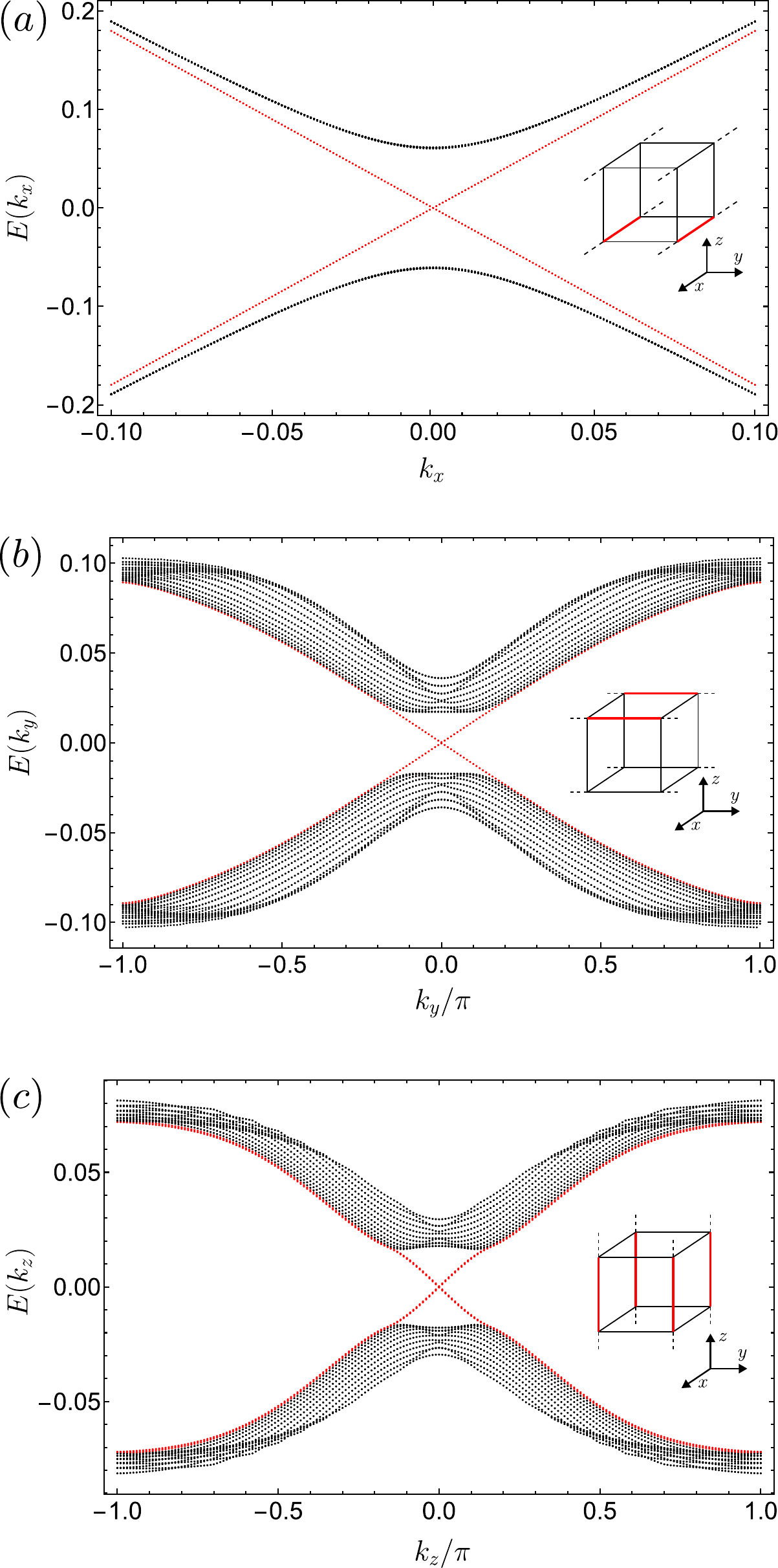}
    \caption{ Numerically calculated low-energy spectra (in units of $E_{\rm so}$) for system parameters as in Fig.~\ref{fig:density}(a) as functions of momentum $k_x$, $k_y$, and $k_z$ for a system periodic along $x$, $y$, and $z$, respectively. Gapless modes (red) appear in all three cases. 
    Both right and left moving edge modes  in the $x$ and $y$ directions are doubly degenerate due to time-reversal symmetry, while the modes propagating along the $z$ direction are fourfold degenerate, as hinge modes are present at all four hinges [see also Fig.~\ref{fig:density}(a)]. 
    }
    \label{fig:surface_spectra}
\end{figure}

From the exact diagonalization results, we may also extract the low energy excitation spectra, as shown in Fig.~\ref{fig:surface_spectra}. 
We first consider the system to be periodic in the $x$ direction and finite in the $y$ and $z$ directions, and in Fig.~\ref{fig:surface_spectra}(a), we plot the energy of lowest energy eigenstates as function of $k_x$ for the set of parameters corresponding to Fig.~\ref{fig:density}(a). In addition to numerous states (black) separated by a surface gap, we find a Kramers pair of gapless hinge modes (red) on a given hinge. All points on this plot are thus doubly degenerate.  Similarly, in Figs.~\ref{fig:surface_spectra}(b) and~\ref{fig:surface_spectra}(c), we show the spectrum as a function of $k_y$ and $k_z$. As expected [see Fig.~\ref{fig:layer_stack}(a)], we find two and four hinge modes (and their Kramers partners) crossing the surface gap. 
The gap sizes in Fig.~\ref{fig:surface_spectra} and localization lengths in Fig.~\ref{fig:density} differ across the different panels because different processes are responsible for gapping out the respective surfaces.

\section{Parafermionic Hinge Modes}\label{sec:Parafermion}
We now turn to the question of whether interactions can yield even more exotic hinge modes. In the previous section, we considered chemical potential $\mu = 0$ and found that hopping could gap out the bulk and surface modes, leaving helical Majorana hinge modes. To realize more exotic modes, we need to dress Majorana excitations with an interaction. In this section, we use bosonization to discuss how this can be done to realize parafermionic hinge modes. 

To ensure that the hopping terms in the previous section need to  be dressed with an interaction term to gap out the system, we tune the chemical potential to  the value \begin{equation}\mu=\left (-1+\frac{1}{q^2}\right)E_{\rm so}, \end{equation} where $q$ is an odd integer. The new Fermi momenta are given by $k_F^{r\tau\sigma}=(r/q + \sigma\tau)k_{\rm so}$. For $q>1$, the tunneling term $H_\Gamma$ cannot open gaps alone, since scattering between different Fermi points does not conserve momentum~\footnote{Within the bosonization framework, the scattering is associated with rapidly varying phase factor which suppresses the term.}. A momentum conserving term achieving this can however be constructed by dressing hopping terms with electronic backscattering arising from electron-electron interactions \cite{Kane2014,Giamarchi}. 
In terms of the composite DNW index $w = (m,n,\gamma,\delta)$, this new term is given by 
\begin{align}
   &H^{w, \, \Gamma'}_{\rm DNW} =i \Gamma' \sum_{\kappa}   \int dx\ \bigg[\left(\bar{R}^{\dagger}_{w \kappa
  \bar{\kappa}}\bar{L}_{w \kappa
  \bar{\kappa}}\right)^{p} \nonumber \\ &\times \left(\bar{R}^{\dagger}_{w \kappa
  \bar{\kappa}} \bar{L}_{w \kappa \kappa}\right)\left(\bar{R}^\dagger_{w \kappa \kappa}\bar{L}_{w \kappa \kappa}\right)^p + \text{H.c.}\bigg],
\end{align} 
where $p=(q-1)/2$. The associated coupling constant is $\Gamma' \propto \Gamma g_{\rm B}^{q-1}$, where we assume that the strength $g_{\rm B}$ of a single back-scattering process caused by electron-electron interactions is large \cite{Klinovaja_Parafermion_Bundle,Sagi2017,Oreg_Fractional_Helical_Liquids,Sine_Gordon_1}. We can similarly write a dressed superconducting term 
\begin{align}
& H^{w,\, \Delta'}_{\rm DNW} = \sum_{\kappa,\nu} \delta \kappa\nu \Delta' \int dx\ \bigg[\left(\bar{R}^{\dagger}_{w \kappa 
     \nu} \overline{L }^{\dagger}_{w \kappa  \nu}\right)^p \nonumber \\ & \times \left(\bar{R}^{\dagger}_{w \kappa 
     \nu} \overline{L }^{\dagger}_{w \kappa \bar \nu}\right)\left(\bar{R}^{\dagger}_{w \kappa 
     \bar \nu} \overline{L }^{\dagger}_{w \kappa  \bar \nu}\right)^p + \text{H.c.}\bigg], 
\end{align} where again, $\Delta' \propto \Delta g_B^{q-1}$. In the noninteracting case with $q=1$, these terms trivially reduce to Eq.~\eqref{H_DNW_after_transformation}.
In order to treat this interaction Hamiltonian analytically, we resort to bosonization~\cite{Giamarchi,Chua_Majorana_Bosonization}.
We therefore define the bosonic fields $\phi^r_{w\kappa\nu}$ through
\begin{subequations} \label{bosonizationbar}
\begin{align}
    \bar R_{w\kappa\nu}(x) &=e^{i\phi^{1}_{w\kappa\nu}(x)},
\\
    \bar L_{w\kappa\nu}(x) &=e^{i\phi^{\bar{1}}_{w\kappa\nu}(x)},
\end{align} 
\end{subequations} where we omitted Klein factors and the ultraviolet cutoffs for the sake of simplicity ~\cite{Kane2014}. The bosonic fields obey the standard commutation relation
\begin{align}
    & \big[ \phi^r_{ w\kappa\nu}(x), \,
       \phi^{r'}_{w'\kappa'\nu'}(x') \big]
   = i \pi r \delta_{rr'} \delta_{ww'}  \delta_{\kappa\kappa'} 
     \delta_{\nu\nu'}  \operatorname{sgn}(x - x') .
\end{align} 
Motivated by the form of the dressed superconducting and tunneling terms $H^{w, \, \Delta'}_{\rm DNW}$ and $H^{w,\, \Gamma'}_{\rm DNW}$, we introduce the new fields
\begin{equation}
\eta^r_{w\kappa\nu}(x)=\left(\frac{q+1}{2}\right)\phi^r_{w\kappa\nu}(x)-\left(\frac{q-1}{2}\right)\phi^{\bar{r}}_{w\kappa\nu}(x),
\end{equation} 
obeying the commutation relation \begin{align}
 \big[  \eta^r_{w\kappa\nu}(x), \,
        \eta^{r'}_{ w'\kappa'\nu'}(x') \big] 
   =  i \pi qr \delta_{rr'}\,
     \delta_{ww'} \delta_{\kappa\kappa'}
     \delta_{\nu\nu'} \operatorname{sgn}(x - x').
\end{align}
The terms responsible for the opening of gaps in the spectrum of the double nanowire Hamiltonian take the form 
\begin{align}
&H_{\rm DNW, I}^{w}= 2   \sum_{\kappa}  \int  dx\  \Big[\Gamma'\sin(\eta^1_{w\kappa\bar\kappa}-   \eta^{\bar{1}}_{w\kappa\kappa})  
     \\ 
   &+\Delta'\cos \left(\eta^1_{w\kappa\bar\kappa}+\eta^{\bar{1}}_{w\kappa\kappa}\right)
   +\Delta' \cos \left(\eta^1_{w\kappa\kappa}+\eta^{\bar{1}}_{w\kappa\bar\kappa} \right) \Big]. \nonumber
\end{align}  
We again see that the exterior branches are fully gapped by the third term in $H_{\rm DNW, I}^{w}$. The first and second terms compete to gap out the interior modes. 
Focusing only on the interior modes of the above DNW Hamiltonian and introducing the new fields 
\begin{align}
\Theta_{w \kappa} & = \frac{
{\eta^1_{w\kappa\bar\kappa}+\eta^{\bar{1}}_{ w \kappa\kappa}}}{2\sqrt{q}},
\\
\Phi_{w \kappa} &= \frac{
\eta^1_{w\kappa\bar\kappa}-\eta^{\bar{1}}_{w \kappa\kappa}+\pi/2}{2\sqrt{q}},
\end{align}
we arrive at 
\begin{align}
H^{w}_\mathrm{DNW, I} &= 2  \sum_{\kappa}\int dx \bigl[ \Gamma' \cos \left(2\sqrt q\Phi_{w\kappa} \right) \nonumber \\ & + \Delta' \cos \left( 2\sqrt q\Theta_{w\kappa}\right)  \bigr].
\label{DNW_Sine_Gordon}
\end{align} 

At the special surface $\Gamma'= \Delta'$ in parameter space, this Hamiltonian corresponds to two time-reversed copies of a self-dual sine-Gordon model \cite{Sine_Gordon_1,Sine_Gordon_2}. For $q=1$, one expects to find two counterpropagating Majorana modes per time-reversal sector, consistent with what we saw in the previous section. In the regime that we are interested in, the competing terms in the DNW Hamiltonian (\ref{DNW_Sine_Gordon}) have the same scaling dimension, allowing us to study the system along the self-dual line \cite{Klinovaja_Parafermion_Bundle,Sagi2017,Oreg_Fractional_Helical_Liquids,Sine_Gordon_1,Sine_Gordon_2,PhysRevB.103.235410}. 

The fixed point of this model corresponds to a gapless phase described by a $\mathbb Z_{2q}$ parafermion theory \cite{Sine_Gordon_2}, so that we have two Kramers pairs of counterpropagating $\mathbb Z_{2q}$ parafermions in each of the DNWs.

Let us now refermionize the above model to obtain the field operators of this parafermionic theory. We define
\begin{equation}
    \bar \psi^{(q)}_{w\kappa\nu}(x)=\bar R^{(q)}_{w\kappa\nu}(x)e^{iq_F^{1\kappa\nu} x}+\bar L^{(q)}_{w\kappa\nu}(x)e^{iq_F^{\bar 1\kappa\nu} x},
\end{equation} where the new Fermi momenta are given by~\cite{Sagi2014} 
\begin{equation}q_F^{r\kappa\nu}=\frac{q+1}{2}k_F^{r\kappa\nu}-\frac{q-1}{2}k_F^{\bar r \kappa \nu}=k_{\rm so}(\kappa\nu+r),\end{equation} and the composite chiral operators $\bar R^{(q)}_{w\kappa\nu}$ and $\bar L^{(q)}_{w\kappa\nu}$ are defined as
\begin{subequations} \label{refermionization}
\begin{align}
\bar R^{(q)}_{w\kappa\nu} &= e^{i\eta^{1}_{w\kappa\nu}}, \\ 
\bar L^{(q)}_{w\kappa\nu} &= e^{i\eta^{\bar 1}_{w\kappa\nu}}.
\end{align}
\end{subequations}

For the interior branches, the tunneling term $H_{\Gamma'}$ and the superconducting term $H_{\Delta'}$  are now given by 
\begin{align}
    H_{\rm DNW}^{w,\, \Gamma'}&=i\Gamma' \sum_\kappa \int dx\ \bar R^{(q)\dagger}_{w\kappa\bar\kappa}\bar L^{(q)}_{w\kappa\kappa}+\text{ H.c.},
\\
    H_{\rm DNW}^{w,\, \Delta'}&=-\delta\Delta' \sum_\kappa \int dx\ \bar R^{(q)\dagger}_{w\kappa\bar\kappa}\bar L^{(q)\dagger}_{w\kappa\kappa}+\text{ H.c.},
\end{align} 
which up to the redefinition of the operators is exactly the same expression as in the noninteracting case. 

As before, we may now introduce the Hermitian fields $\chi^{R/L (q)}_{w \kappa\nu}$ and $\bar{\chi}^{R/L (q)}_{w \kappa\nu}$ through
\begin{subequations}
\begin{align}
\bar{R}^{(q)}_{w\kappa\nu} &= \frac{e^{i \pi /
    4}}{\sqrt{2}} (\chi^{(q)R}_{w\kappa\nu} + i \bar\chi^{(q)R}_{w\kappa\nu}),
\\
\bar{L}^{(q)}_{w\kappa\nu} &= \frac{e^{i \pi /
    4}}{\sqrt{2}} (\chi^{(q)L}_{w\kappa\nu} + i \bar\chi^{(q)L}_{w\kappa\nu}).
\end{align} 
\end{subequations}
By the same steps  as in the noninteracting case, one can then show that the modes 
\begin{subequations}
\begin{align}
\chi^{(q) \kappa R }_w&=\begin{cases}\chi^{(q)R}_{w\kappa\bar\kappa} & \text{if }\delta=1 \\ \bar \chi^{(q)R}_{w \kappa\bar\kappa}  & \text{if } \delta=\bar 1\end{cases}\ ,
\\
\chi^{(q) \kappa L}_{w}&=\begin{cases}\chi^{(q)L}_{w\kappa\kappa} & \text{if }\delta=1 \\ \bar \chi^{(q)L}_{w \kappa\kappa}  & \text{if } \delta=\bar 1\end{cases}\ , 
\end{align} 
\end{subequations}
are left gapless, where we recall that the index $\delta$ is contained in the composite index $w$. These Hermitian fields can be identified as the primary fields of the $\mathbb Z_{2q}$ parafermionic theories describing each DNW. 

Analogous to the dressed tunneling and superconducting contributions introduced above for the DNW Hamiltonian, we can construct the dressed versions of the remaining terms of the full Hamiltonian in Eq.~\eqref{eq:Htotal}, which we call $\tilde H_{z}'$, $H_z'$, $\tilde H_y'$, and $H_y'$. An example of how to construct such terms is given in App.~\ref{appendix:dressed_terms_appendix}, where we provide details on the derivation of $\tilde H_z'$. In terms of parafermionic field operators, we find 
\begin{subequations}
\begin{align}
\tilde H'_{z} &= \frac{i}{2} \sum_{m,n,\delta,\kappa} \int dx \, \Big[
      (\kappa\beta' - \Delta_c') 
\chi^{(q)\kappa R}_{mn \bar 1 \delta} \chi^{(q)\kappa L}_{mn1 \delta}
      \nonumber \\
 & +  (\kappa\beta' + \Delta_c') 
\chi^{(q)\kappa L}_{mn \bar 1 \delta} \chi^{(q)\kappa R}_{mn1 \delta}
 \Big], \label{intra_Hz_as_parafermion} \\
H_{z}' &= \frac{i}{2} \sum_{m,\delta,\kappa} \sum_{n=1}^{N_z-1} \int dx \, \Big[
    (\kappa\beta' - \Delta_c') 
\chi^{(q)\kappa R}_{m(n+1)1\delta} \chi^{(q)\kappa L}_{mn\bar 1\delta} \label{inter_Hz_as_parafermion}
    \nonumber \\
&+ (\kappa\beta' + \Delta_c') 
\chi^{(q)\kappa L}_{m(n+1)1\delta} \chi^{(q)\kappa R}_{mn\bar 1\delta}
\Big],
\end{align}
\end{subequations}
\begin{subequations}\label{eq:hy_parafermion}
\begin{align} \label{finalformintracell_parafermion}
\tilde H_{y}' &= \frac{i}{2} 
    \sum_{n\in \{1, N_z\}}
    \sum_{m,\gamma,\kappa} \kappa \, \tilde t_y^{\gamma'} \int dx \,   \Big[
   \chi^{(q)\bar{\kappa} R}_{mn\gamma\bar{1}}\chi^{(q)\kappa L}_{mn\gamma 1}
    \\ \nonumber 
&\quad + 
\chi^{(q)\bar\kappa L}_{mn\gamma\bar1}\chi^{(q)\kappa R}_{mn\gamma1}
\Big], \\
\label{ycouplingfinalintercell_parafermion}
H_y' &= \frac{i}{2} 
    \sum_{n\in \{1, N_z\}}
    \sum_{m=1}^{N_y-1} 
    \sum_{\gamma,\kappa} \kappa \,
    t_y^{\gamma'} \int dx \, \Big[
    \chi^{(q)\kappa R}_{(m+1)n\gamma 1}  {\chi}^{(q)\bar \kappa L}_{mn\gamma\bar 1} 
    \nonumber \\
&\quad + 
\chi^{(q)\kappa L}_{(m+1)n\gamma 1}\chi^{(q)\bar \kappa R}_{mn\gamma\bar 1} 
\Big].
\end{align}
\end{subequations} Here, we assumed the hierarchy \begin{equation}
E_{\rm so} \gg \Delta',\, \Gamma' \gg \beta',\, \Delta_c' \gg t_{y}^{1'},\ 
\tilde{t}_{y}^{\,\bar{1}'} \gg \tilde{t}_{y}^{1'},\ t_{y}^{\bar{1'}} \ge 0,
\end{equation}
where each primed amplitude scales as its noninteracting (unprimed) counterpart multiplied by $g_{\rm B}^{\,q-1}$. As can be checked, these Hamiltonian terms are identical to those we found in the noninteracting case [see Eqs.~\eqref{z_terms_majorana}--\eqref{eq:hy}], up to the replacement $\chi^{\kappa (R/L)}_{mn\gamma\delta} \rightarrow \chi^{(q) \kappa (R/L)}_{mn\gamma\delta}$. Provided that the dressed terms remain RG-relevant or are initially strong, both the bulk and the surfaces of the 3D array of Rashba nanowires are thus fully gapped. Importantly, the dressed interacting Hamiltonian is also block-diagonal in the time-reversal sector $\kappa$. Our entire low-energy effective theory of noninteracting electrons at $\mu=0$ hosting Majorana hinge modes is therefore identical to that of interacting electrons at $\mu=\left(-1+\frac{1}{q^2}\right)E_{\rm so}$~\cite{Kane2014,Kane2002,Laubscher2023,Sagi2015b,Sagi2017}. Since the primary fields of the interacting theory are the parafermion operators $\chi^{(q) \kappa (R/L)}_{mn\gamma\delta}$, we thus find a Kramers pair of zero-energy $\mathbb Z_{2q}$ parafermionic hinge modes propagating along a closed path of 1D hinges of the sample. Just like the Majorana modes of Sec.~\ref{sec:Majorana}, these parafermionic modes are topologically protected by both particle-hole and time-reversal symmetry, and their precise geometry depends on the dimerization pattern fixed by the relative size of the hoppings $\tilde t_y^{\,\bar 1'}$ and $\tilde t_y^{1'}$ [see Subsec.~\ref{subsec:yzSurface} and App.~\ref{appendix:finite_nanowire_states}], and on the boundary termination of the sample [see Fig.~\ref{fig:layer_stack}].

\section{CONCLUSIONS AND OUTLOOK}\label{sec:conclusion}

In this work, we developed a model for a 3D second-order topological superconductor using an array of weakly coupled superconducting Rashba nanowires. We showed that in a certain region of parameter space, depending on the Fermi energy $\mu$, the system supports Majorana hinge states or, in the presence of strong electron-electron interactions, more general $\mathbb{Z}_{2q}$ parafermionic hinge states, with $q$ being an odd integer. Both these types of quasiparticles exhibit non-Abelian braiding statistics~\cite{Kane2002,Kane2014,Beenakker2019Search,Schmidt2019Bosonization,Ivanov2000Non-Abelian}. The paths taken by these hinge modes can be understood by mapping the system onto an effective SSH model, where the relative strength of intra- and intercell hoppings, together with the boundary termination, dictates whether the system realizes a trivial or a non-trivial higher-order topological phase. We also investigated the stability of these hinge modes away from special fine-tuned points in parameter space which simplified the analytical analysis, and with respect to random onsite charge disorder of varying strength. In both cases, we found the hinge states to be remarkably robust: although our analytical treatment relies on a multi-step perturbation procedure, the existence of these modes does not really depend on perturbative gaps. In fact, they persist as long as both the bulk and the two-dimensional surfaces remain gapped.

The coupled-wire framework employed in this work should be viewed as a minimal theoretical model. A direct realization of the full 3D array with its specific parameter hierarchies and staggering would be experimentally challenging because of uniformity, disorder, and precise control. Nevertheless, the model is motivated by substantial experimental progress in proximitized semiconductor nanowires and networks. One-dimensional Rashba nanowires with proximity-induced superconductivity have been extensively investigated and have produced signatures consistent with Majorana zero modes \cite{Lutchyn2018}. Building on this, proposals exist for 2D arrays of such nanowires that could host tunable second-order topological superconductivity \cite{Franca2019}, with inter-wire couplings adjustable, for example, through phase biasing. Key experimental parameters, such as gate-tunable Rashba SOI and chemical potential, are already accessible in these systems. While uniform 3D nanowire arrays remain technologically demanding, advances in selective-area growth of complex nanowire networks \cite{Yuan2021}, multilayer heterostructures, and alternative platforms (such as magnet-superconductor hybrids \cite{Wong2023}) offer possible routes for related implementations in the future. To this end, the robustness of the predicted hinge modes to moderate disorder is especially encouraging. Thus, we believe that despite the experimental challenges and the number of microscopic parameters in our model, its underlying physical mechanism is simple and provides a transparent and broadly applicable framework for engineering higher-order topological superconductivity.

\acknowledgments
We would like to thank Valerii Kozin, Maximilian Hünenberger, Katharina Laubscher, and Peter Daniel Johannsen for fruitful discussions. This work was supported by the Swiss National Science Foundation and NCCR SPIN (grant no. 51NF40-180604).

\section*{Data Availability}
The data corresponding to the findings of this article are openly available \cite{data}.

\appendix

\section{Geometry of hinge modes in finite nanowires} \label{appendix:finite_nanowire_states}

In this Appendix, we derive the wave functions corresponding to the bilayers labeled by the indices $(m, \delta)$, as introduced in the main text. For convenience, we work in the $(\kappa, \nu)$ basis rather than the $(\tau, \sigma)$ basis [see Eq.~\eqref{eq:unitary_transformation}]. The wave functions in this basis can be obtained straightforwardly in momentum space. 

To capture finite-size effects along the $x$ direction, we assume that the system is periodic along the $z$ direction, which allows us to perform a Fourier transform to characterize the states by the momentum $k_z$. Since the intra- and inter-cell couplings along the $z$ axis are identical, it is not necessary to explicitly include the index $\gamma$ in the field operator~\cite{Note2} as long as we focus on properties of bilayers. Consequently, each unit cell along the $z$ axis effectively consists of a single nanowire. The corresponding Fourier-transformed field operator is defined as
\begin{equation}
\bar{\psi}_{m k_z \delta \kappa \nu}(x) = \frac{1}{\sqrt{N_z}} \sum_{n} e^{i n k_z a_z} \, \bar{\psi}_{m n \delta \kappa \nu}(x),
\end{equation}
where $a_z$ denotes the lattice constant along the $z$ direction, which we set to unity from now on.

The Hamiltonian of bilayer $\delta$ in the $m^{\text{th}}$ unit cell is then given by
\begin{align} \label{bilayerBdGwithoutlinearisation}
    H_{m\delta}& =\sum_{k_z}\int dx\, \bar{\Psi}^\dagger_{mk_z\delta}\bigg [ \left(-\frac{\partial_x^2}{2m_e}-\mu \right)\eta_z \nonumber \\&  + i\alpha\kappa_z\nu_z \partial_x+ \beta\sin(k_z)\kappa_z\nu_x + \Gamma\kappa_z\nu_y \nonumber  \\& + \delta(\Delta+\Delta_c\cos k_z)\kappa_z\eta_y\nu_y  \bigg]\bar\Psi_{mk_z\delta},
\end{align}
which is diagonal in momentum $k_z$, and $\bar\Psi_{m k_z \delta}$ is the spinor
\begin{align} \label{kappa_eta_nu_basis}\bar\Psi_{m k_z \delta}=& \big( \bar{\psi}_{m k_z\delta 11 },\  \bar{\psi}_{m k_z \delta 1\bar 1 },\ \bar \psi^\dagger_{m -k_z \delta 11 },\ \bar\psi^\dagger_{m - k_z \delta 1\bar 1, } 
\nonumber\\&\ 
\bar\psi_{m - k_z \delta \bar 1 1 },\ \bar\psi_{m - k_z \delta \bar 1 \bar 1 },\ \bar \psi^\dagger_{m - k_z \delta \bar 1 1},\ \bar\psi^\dagger_{m - k_z \delta  \bar 1 \bar 1}\big)^T \end{align} 
Again, the operator $\bar \psi_{m k_z  \delta \kappa \nu}(x)$ can be represented in terms of slowly varying right- and left-moving fields $\bar R_{m k_z \delta \kappa \nu}(x)$ and $\bar L_{m k_z \delta \kappa  \nu}(x)$, defined close to the Fermi points $k_F^{r\kappa\nu}$, via
\begin{equation}
\bar \psi_{m k_z \delta \kappa \nu} = e^{i k_F^{1\kappa\nu} x} \bar R_{m k_z \delta \kappa \nu} + e^{i k_F^{\bar 1\kappa\nu} x} \bar L_{m k_z \delta \kappa \nu}.
\end{equation}
It can be checked that for any $m$ and $\delta$, for all our choices of parameters [see Table~\ref{tab:parameters}], the corresponding Bogoliubov-de-Gennes (BdG) Hamiltonian in Eq.~\eqref{bilayerBdGwithoutlinearisation} gives zero energy Majorana modes at the time-reversal invariant momentum $k_z= \pi$. Therefore, in order to find the gap-opening effects of the couplings $\tilde t_y^{\, \bar 1}$ and $t_y^1$ on $yz$ surfaces, it is sufficient to look at the corresponding overlap integrals of these zero-energy wave functions at the point $k_z= \pi$ in momentum space. We define the Hamiltonian density $\mathcal{H}_{m k_z\delta}$ through $H_{m k_z \delta} = \int d x \ \bar \Phi_{mk_z\delta}^{\dagger}(x) \mathcal{H}_{mk_z\delta} \bar \Phi_{mk_z\delta}(x)$. This Hamiltonian density is
\begin{align} &\mathcal H_{mk_z\delta}=-(i\alpha\partial_x)\lambda_z +\frac{\beta}{2} \sin k_z \left(\kappa_z\nu_x\lambda_x+\nu_y\lambda_y\right)\label{eq:H_m_delta} \nonumber \\& +\frac{\Gamma}{2}(\kappa_z\nu_y\lambda_x-\nu_x\lambda_y)+\delta \left(\Delta+\Delta_c\cos k_z\right)\kappa_z\eta_y\nu_y\lambda_x \end{align} 
in terms of the basis $(\kappa\otimes\eta\otimes\nu \otimes\lambda)$
\begin{align}
\bar \Phi_{mk_z\delta} &=
\Big(
\bar R_{mk_z11},\ \bar L_{mk_z11},\ \bar R_{mk_z1\bar 1},\ \bar L_{mk_z1\bar 1}, \nonumber \\
&
\bar R^\dagger_{m-k_z11},\  \bar L^\dagger_{m-k_z11}, \ \bar R^\dagger_{m-k_z1\bar 1},\ \bar L^\dagger_{m-k_z1\bar 1}, \nonumber \\& \bar R_{mk_z\bar 1 1},\ \bar L_{mk_z\bar 1 1},  \bar R_{mk_z\bar 1 \bar 1},\ \bar L_{mk_z\bar 1 \bar 1}, \nonumber \\ &
\bar R^\dagger_{m-k_z\bar 1 1},\ \bar L^\dagger_{m-k_z\bar 1 1},\ \bar R^\dagger_{m-k_z\bar 1 \bar 1},  \bar L^\dagger_{m-k_z\bar 1 \bar 1} 
\Big)^{T},
\end{align} 
where $\lambda \in \{R,L\}$, and the Pauli matrices $\kappa,\ \eta,\ \nu,\ \lambda$ act in the pseudolayer, particle-hole, pseudospin and left/right mover space respectively. Considering the nanowires to be semi-infinite, we then impose vanishing boundary conditions at the left end of each wire, i.e. we require $\bar \Psi_{mk_z\delta}(x = 0) = 0$ and $E = 0$ when $k_z = \pi$, where $\bar \Psi_{mk_z\delta}(x )$ is the eigenfunction of the operator $\mathcal{H}_{mk_z \delta}$. For each $\delta$, this yields two degenerate solutions labeled by the indices $\kappa=1$ and $\kappa=\bar 1$. These solutions are written in  basis $\kappa \otimes \eta \otimes \nu$ [see Eq. \eqref{kappa_eta_nu_basis}] as
\begin{subequations} \label{edge_state_Majorana} 
\begin{align} \big|\bar{\Psi}^{\kappa=1, \pi}_{m \delta}
    \big\rangle&=  \frac{1}{\sqrt{N}} \big( e^{i\delta\pi/4}f^\ast,\  -e^{i\delta\pi/4}f,  \quad  e^{-i\delta\pi/4}f, \nonumber \\& \quad  -e^{-i\delta\pi/4} f^\ast,\  0, \ 0,\ 0,\ 0  \big)^T, 
\\
\big|\bar{\Psi}^{\kappa=\bar{1}, \pi}_{m \delta}
    \big\rangle&=  \frac{1}{\sqrt{N}} \big(0, \ 0,\ 0,\ 0,\  e^{i\delta\pi/4}f,\  -e^{i\delta\pi/4}f^\ast, \nonumber \\& \quad  e^{-i\delta\pi/4}f^\ast,  \quad  -e^{-i\delta\pi/4} f \big)^T,\end{align} 
\end{subequations}    
where $ f(x)=e^{-2ik_{\rm so}x}e^{-x/\xi_2}-e^{-x/\xi_1}$ and $N$ 
    is the real normalization factor
\begin{equation}
    N=\sqrt{2(\xi_1+\xi_2) \frac{(\xi_1 - \xi_2)^2 + 4 k_\mathrm{so}^2 \xi_1^2 \xi_2^2 }{(\xi_1 + \xi_2)^2 + 4 k_\mathrm{so}^2 \xi_1^2 \xi_2^2} },
\end{equation}
where we introduced localization lengths $\xi_1 = \alpha/\Delta_c$, $\xi_2 = \alpha/(\Delta-\Delta_c)$. The two wave functions $\big|\bar{\Psi}^{\kappa=1, \pi}_{m \delta} \big\rangle$ and $\big|\bar{\Psi}^{\kappa=\bar{1}, \pi}_{m \delta} \big\rangle$ form a time-reversal pair, and each satisfies the Majorana condition: its particle and hole components are related by complex conjugation, as dictated by the structure of the BdG Hamiltonian.

Next, we calculate the relevant matrix elements between these bilayer states to find under which condition the couplings along the $y$ axis opens gaps on the $yz$ surface at $x=0$. In real space, the relevant matrix connecting states in neighboring bilayers along the $y$ direction is $\left[(\delta_x\tau_x+\delta_y\tau_y)\eta_z\sigma_0 \right]/2$ [see Eqs. (\ref{yintracellnewcoupling}) and (\ref{intercellnewcoupling})], which, upon transforming to $(\kappa,\nu)$ space becomes $\left(\delta_x\kappa_z\eta_z\nu_y+\delta_y\kappa_y\eta_z\nu_0\right)/2$. Since the first of these terms conserves the $\kappa$ index, the corresponding matrix element is zero, and the term does not contribute to surface gaps. The second term flips the $\kappa$ index and therefore gives an effective coupling of the form 
\begin{subequations}
\begin{align}
\left\| \left\langle \bar{\Psi}^{\kappa \pi}_{m1} \Big| \tilde{H}_y \Big| \bar{\Psi}^{\bar\kappa \pi}_{m \bar{1}}\right\rangle  \right\|&= \frac{\tilde t_y^{\, \bar 1}}{2} \left\| \left\langle \bar{\Psi}^{\kappa \pi}_{m 1} \Big| \kappa_+\eta_z\nu_0\Big| \bar{\Psi}^{\bar{\kappa}\pi}_{m \bar{1}} \right\rangle  \right\|,
\\
\label{appendix_intercell_y_overlap}
    \left\| \left\langle 
    \bar{\Psi}^{\bar{\kappa} \pi}_{m\bar{1}} \Big| H_y \Big| 
    \bar{\Psi}^{\kappa \pi}_{(m+1) 1 }
    \right\rangle \right\| 
    &= \frac{t_y^{1}}{2} \left\| \left\langle \bar{\Psi}^{\bar{\kappa}\pi}_{m \bar 1 } \Big|  \kappa_+ \eta_z\nu_0\Big| \bar{\Psi}^{\kappa \pi}_{(m+1) 1} \right\rangle  \right\|,
\end{align}
\end{subequations}
where $\kappa_+ = \kappa_1 + i \kappa_2$. Exploiting the explicit form of the eigenstates, we therefore obtain 
\begin{subequations}
\begin{align}
&\left\| \left\langle \bar{\Psi}^{\kappa \pi}_{m1} \Big| \tilde{H}_y \Big| \bar{\Psi}^{\bar\kappa \pi}_{m \bar{1}}\right\rangle  \right\| = c\ \tilde t_y^{\, \bar 1}, \\
&\left\| \left\langle 
    \bar{\Psi}^{\bar{\kappa} \pi}_{m\bar{1}} \Big| H_y \Big| 
    \bar{\Psi}^{\kappa \pi}_{(m+1) 1 }
    \right\rangle \right\| = c\ t_y^{1},
\end{align}   
\end{subequations}
where the common positive constant is
\begin{align}
c = \frac{2}{N^2} \int_{0}^{\infty}dx\ ([f(x)]^2+[f^\ast(x)]^2).  
\end{align}
When the couplings $\tilde t_y^{\, \bar 1}$ and $t_y^1$ are equal, the $yz$ surfaces of the 3D sample are left gapless, as discussed in more detail in the main text. On the other hand, when amplitudes differ, the dimerization pattern decides the geometry of the resultant hinge modes at the $yz$ surfaces.

Although the unit cell of the full system contains two nanowires along the $z$-direction, in this Appendix we considered a single bilayer $(m,\delta)$ and therefore retain only one nanowire per unit cell along $z$. Consequently, the Majorana modes found at $k_z = \pm \pi $ in this reduced description are folded back to $k_z=0$ in the full model used for the numerical calculations shown in Fig.~\ref{fig:surface_spectra}(c).

\section{Details of the tight-binding simulation} 
\label{appendix:tight_binding_details}
In this appendix, we provide a discussion of how the numerical results were obtained.
Assuming that the number of sites in the $x$-direction is large,
we can perform a faithful discretization of the Hamiltonian with the tight-binding approximation. Given that we have eight nanowires per unit cell, along with particle-hole space and spin space, we get an internal Hilbert space of dimension $2^5=32$. Introducing the electron annihilation operator $\psi^l_{mn\gamma\delta\tau\sigma}$, where $l$ is the lattice site in the $x$ direction and all other symbols have the meaning introduced in the main text, we find
\begin{equation}\label{eq:HtotalAppendix}
H = H_0 +H_\Delta + H_{\Gamma}+ \tilde H_z + H_z + \tilde H_y + H_y,
\end{equation} 
where the individual terms are given by
\begin{align}\label{eq:kineticAppendix}
H_{0} &= \sum_{l,m,n} \sum_{\delta, \tau, \gamma, \sigma} 
    \Big\{(2t_x-\mu)\left(\psi^{l}_{mn\gamma\delta\tau\sigma}\right)^{\dagger}\psi^{l}_{mn\gamma\delta\tau\sigma} \nonumber \\ &
    -\left[(t_x-i\tau\sigma\tilde \alpha)\left(\psi^{l}_{mn\gamma\delta\tau\sigma}\right)^\dagger\psi^{l+1}_{mn\gamma\delta\tau\sigma}+ \text{H.c.} \right]\Big\},
 \\
\label{proximityinducedsuperconductivityAppendix} H_{\Delta} &=\Delta \sum_{l,m,n}\sum_{\delta,\tau,\gamma} (\delta \tau) \psi^{l}_{mn\gamma\delta\tau 1} \psi^{l}_{mn
\gamma\delta\tau\bar{1}}+ \text{H.c.},
\\
\tilde H_{z} &=  \frac{1}{2}  \sum_{\substack{l,m,n,\\ \delta, \tau,\sigma, \sigma' }}   \Big[ i \delta\tau \Delta_c
     \psi^{l}_{mn 1 \delta \tau \sigma}  \sigma_y^{\sigma \sigma'}   \psi^{l}_{mn\bar 1\delta\tau \sigma'}  \nonumber\\ &-  i\beta \tau \left(\psi^{l}_{mn 1 \delta \tau \sigma}    \right)^\dagger  \sigma_x^{\sigma \sigma'}\psi^{l}_{mn
     \bar 1\delta \tau \sigma'}+\ \text{H.c.} \Big],\label{intracellHzAppendix} \\ H_{z}&= \frac{1}{2}  \sum_{\substack{l,m,n,\\ \delta, \tau,\sigma, \sigma' }} 
\Big[i \delta\tau \Delta_c
\psi^{l}_{mn \bar 1 \delta \tau \sigma} { \sigma_y^{\sigma \sigma'}}
\psi^{l}_{m(n + 1)1\delta\tau \sigma'} \nonumber \\ 
&- i \beta \tau \left(\psi^{l}_{mn \bar 1 \delta \tau \sigma}\right)^\dagger 
 \sigma_x^{\sigma \sigma'} 
\psi^{l}_{m(n + 1) 1\delta \tau \sigma'} 
+  \text{H.c.} \Big],
\label{intercellHzAppendix}\\ 
\label{yintracellgammacouplingAppendix}
H_{\Gamma} &= \Gamma \sum_{\substack{l,m,n, \\ \delta,\gamma,\sigma}}  
\left(\psi^{l}_{mn\gamma\delta 1 \sigma}\right)^\dagger\psi^{l}_{mn\gamma\delta\bar{1}\sigma} + \text{H.c.},
\\ \label{yintracellnewcouplingAppendix}
\tilde{H}_{y} &= \sum_{\substack{l,m,n, \\ \gamma,\sigma}} \tilde{t}_y^{\gamma} 
\left(\psi^{l}_{mn\gamma 1 \bar{1} \sigma}\right)^\dagger\psi^{l}_{mn\gamma \bar{1} 1 \sigma} + \text{H.c.},
\\
\label{intercellnewcouplingAppendix}
H_{y} &= \sum_{\substack{l,m,n,  \gamma,\sigma}} t_{y}^{\gamma} 
\left(\psi^{l}_{mn\gamma \bar{1} \bar{1} \sigma}\right)^\dagger
\psi^{l}_{(m+1)n\gamma 1 1 \sigma} + \text{H.c.}.
\end{align}
Here, $t_x=1/(2m_{e}a_x^2)$ is the hopping matrix element in the $x$ direction and is equal to $0.9E_{\rm so}$. Further, $\tilde \alpha \equiv \alpha/2a_x=0.95E_{\rm so}.$ 

\section{Hinge modes away from $\mu=0$} in the presence of onsite disorder \label{appendix:disorder}
In this Appendix, we examine how the probability density of hinge modes [see Fig.~\ref{fig:density}(a)] changes in the presence of random onsite charge disorder away from the fine-tuned conditions $\Gamma=\Delta$, $\beta=\Delta_c$ and $\mu=0$. The onsite chemical potentials $\mu^l_{mn\gamma\delta\tau}=\mu+\delta\mu^{l}_{mn\gamma\delta\tau}$ are drawn from a Gaussian distribution with mean $\mu$ and standard deviations (SD) specified in Table ~\ref{tab:parametersII}.

\begin{figure*}[]
    \centering
\includegraphics[width=0.75\textwidth]{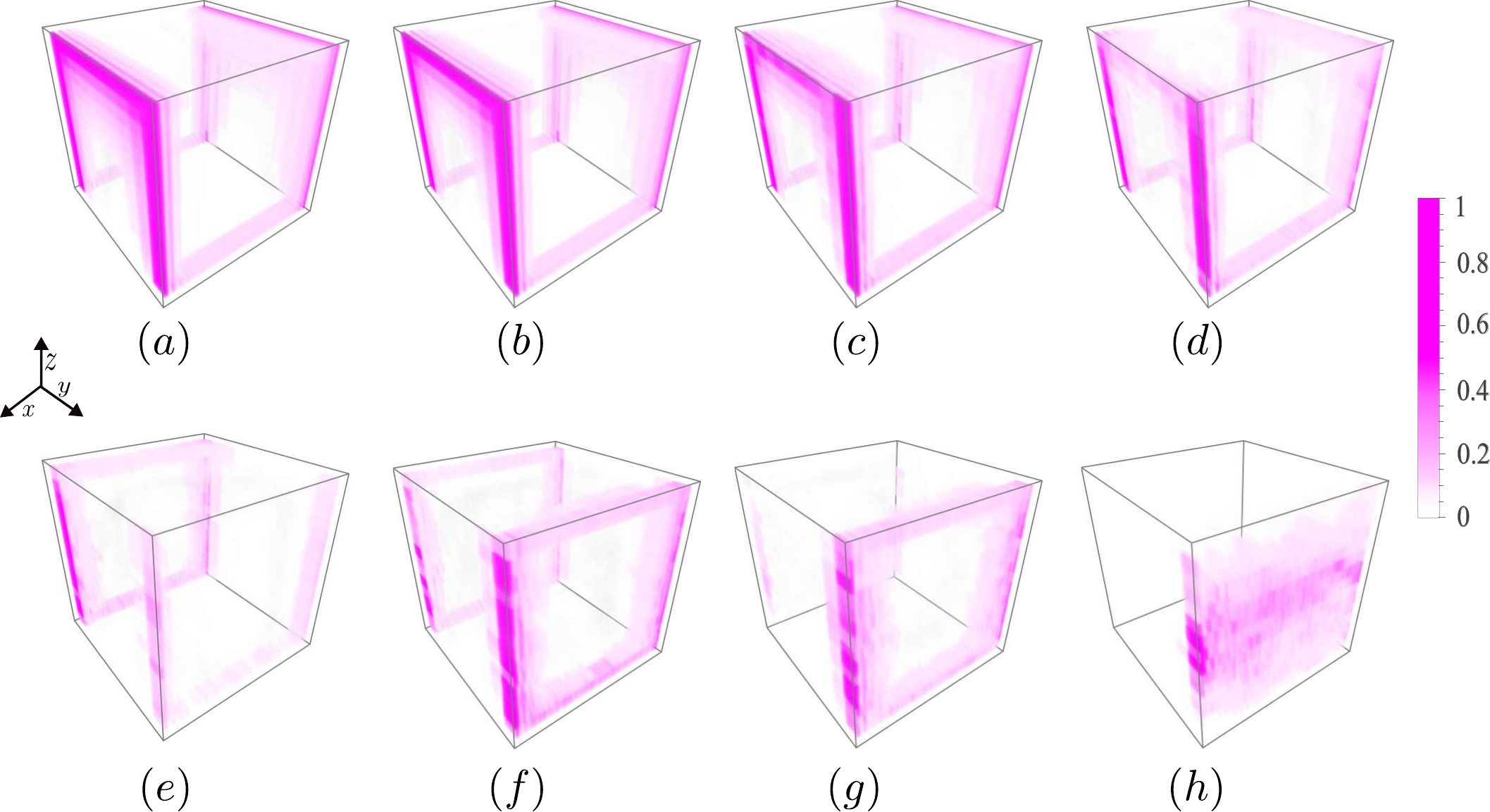}
    \caption{Majorana hinge modes for the geometry in Fig. \ref{fig:density}(a) at $\mu=0$ in the presence of random onsite charge disorder of varying strength away from the fine-tuned points $\Gamma=\Delta$ and $\beta=\Delta_c$ in parameter space. The full set of parameter values is given in Table \ref{tab:parametersII}. (a) Hinge modes at the fine-tuned point in the absence of disorder. (b) Away from the fine-tuned point but without disorder. (c)--(h) Away from the fine-tuned point and with increasing disorder.
    We find that at disorder strength $\text{SD} \approx 0.6 E_{\rm so}$, surface gaps begin to close and the hinge modes disappear. The system has $N_y \times N_z =15 \times 10 $ unit cells with $N_x=100$ sites in the $x$ direction.} \label{fig:density_disorder}
\end{figure*}

First, in Fig.~\ref{fig:density_disorder}, we focus on $\mu=0$ and varying disorder strengths. Evidently, the hinge states are remarkably robust against charge disorder away from the special points $\Gamma=\Delta$ and $\beta=\Delta_c$, up to disorder strengths as large as $0.6E_{\rm so}$, beyond which the surface gaps begin to close. Notably, the hinge states persist even when the disorder strength exceeds the energy scales $t_y^{1}$, $\tilde{t}_y^{\,\bar{1}}$, $\beta$, and $\Delta_c$. This robustness likely originates from the fact that Majorana modes would still exist in all DNWs in the limit where all of these amplitudes vanish, and being charge-neutral, they are insensitive to onsite charge disorder. However, once the disorder strength becomes comparable to the intra-DNW amplitudes $\Gamma$ and $\Delta$, the Majorana modes are destroyed, the effective low-energy description breaks down, and the hinge states cease to exist. 

\begin{figure*}[]
    \centering
\includegraphics[width=0.75\textwidth]{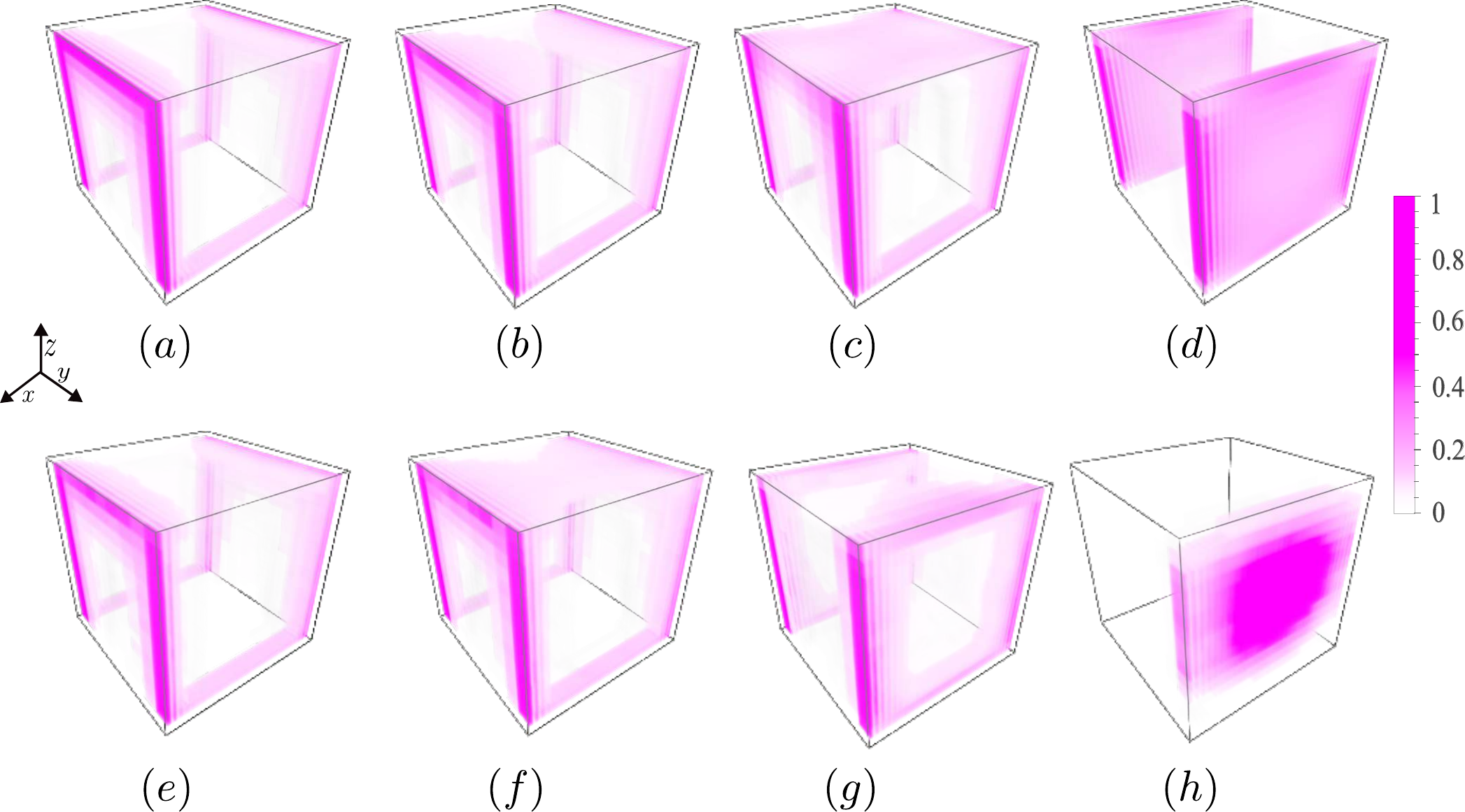}
 \caption{Hinge modes for the geometry in Fig.~\ref{fig:density}(a) away from the point $\mu=0$. The parameter values are given in Table~\ref{tab:parametersII}. (a)--(d) Finite $\mu$ in the clean system. (e)--(h) Finite $\mu$ with fixed nonzero disorder ($\text{SD}=0.15E_{\rm so}$). We find that around $\mu\approx 0.2E_{\rm so}$ the surface gaps begin to close and the hinge modes disappear. This holds both in the clean system and in the presence of weak disorder. The system has $N_y \times N_z =15 \times 10 $ unit cells with $N_x=100$ sites in the $x$ direction.}\label{fig:density_disorder_changing_mu}
\end{figure*}

We further investigate the effect of a finite chemical potential $\mu$ on the Majorana hinge modes, as shown in Fig.~\ref{fig:density_disorder_changing_mu}. The hinge states remain stable up to $\mu\approx 0.2E_{\rm so}$, beyond which the surface gaps close. As discussed in Sec.~\ref{sec:Parafermion}, at a finite $\mu$, the tunneling term $H_\Gamma$ does not conserve momentum. Consequently, $H_\Gamma$ can no longer open gaps at the Fermi surface if the deviation from $\mu=0$ is too large. Due to this, as $\mu$ increases from zero, the Majorana hinge modes first get delocalized and, as the surface gap vanishes, they disappear. 

\begin{table}[h]
\centering
\caption{Parameter values used in the density plots of Figs.~\ref{fig:density_disorder} and \ref{fig:density_disorder_changing_mu}. All energies are given in units of $E_{\rm so}=m_e\alpha^2/2$. The hopping matrix element $t_x=0.9E_{\rm so}$ and $\tilde \alpha=\alpha/2a_x=0.95E_{\rm so}$. The second-last column lists the standard deviation (SD) of the onsite chemical potentials, drawn from a Gaussian distribution with mean $\mu$ listed in the last column.}
\label{tab:parametersII}
\begin{ruledtabular}
\begin{tabular}{ccccccccc}
Plot   & $\Gamma$ & $\Delta$ & $\beta$ & $\Delta_c$ & $t_y^{1}$ & $\tilde{t}_y^{\, \bar{1}}$ & SD & $\mu$ \\ \hline
$8(a)$              & 0.60 & 0.60 & 0.25 & 0.25 & 0.40 & 0.18 & 0 & 0\\
$8 (b)$                & 0.60 & 0.65 & 0.25 & 0.30 & 0.40 & 0.18 & 0 & 0\\
$8(c)$               & 0.60 & 0.65 & 0.25 & 0.30 & 0.40 & 0.18 & 0.15& 0\\
$8(d)$                & 0.60 & 0.65 & 0.25 & 0.30 & 0.40 & 0.18 & 0.45& 0\\
$8(e)$    & 0.60 & 0.65 & 0.25 & 0.30 & 0.40 & 0.18 & 0.55& 0\\
$8(f)$      & 0.60 & 0.65 & 0.25 & 0.30 & 0.40 & 0.18 & 0.60 & 0\\
$8(g)$              & 0.60 & 0.65 & 0.25 & 0.30 & 0.40 & 0.18 & 0.65 & 0\\
$8(h)$              & 0.60 & 0.65 & 0.25 & 0.30 & 0.40 & 0.18 & 0.80 & 0\\
$9(a)$               & 0.60 & 0.65 & 0.25 & 0.30 & 0.40 & 0.18 & 0 & 0.10\\
$9(b)$               & 0.60 & 0.65 & 0.25 & 0.30 & 0.40 & 0.18 & 0 & 0.23\\
$9(c)$               & 0.60 & 0.65 & 0.25 & 0.30 & 0.40 & 0.18 & 0 & 0.30\\
$9(d)$              & 0.60 & 0.65 & 0.25 & 0.30 & 0.40 & 0.18 & 0 & 0.45\\
$9(e)$               & 0.60 & 0.65 & 0.25 & 0.30 & 0.40 & 0.18 & 0.15 & 0.15\\
$9(f)$               & 0.60 & 0.65 & 0.25 & 0.30 & 0.40 & 0.18 & 0.15 & 0.23\\
$9(g)$               & 0.60 & 0.65 & 0.25 & 0.30 & 0.40 & 0.18 & 0.15 & 0.37\\
$9(h)$               & 0.60 & 0.65 & 0.25 & 0.30 & 0.40 & 0.18 & 0.15 & 0.52\\
\end{tabular}
\end{ruledtabular}
\end{table}

\section{Dressed terms for inter-DNW interactions in the 
\texorpdfstring{$y$}{y} and \texorpdfstring{$z$}{z} directions}
\label{appendix:dressed_terms_appendix}

In this Appendix, we discuss how to obtain expressions for the dressed inter-DNW hopping terms. 
Similar to how we wrote down the dressed terms for a double nanowire in the presence of interactions for chemical potential $\mu = (-1+1/q^2)E_{\rm so}$ [see Sec. \ref{sec:Parafermion}], we can write down the dressed terms corresponding to $\tilde H_{z},\ H_{z},\ \tilde H_y,\ H_y$ in the noninteracting Hamiltonian. These dressed terms couple the gapless parafermionic modes in a DNW, and we refer to them as $\tilde H'_{z},\ H'_{z},\ \tilde H'_y,\ H'_y$, respectively. We show the derivation for $\tilde{H}'_z$, while the remaining terms are obtained analogously.
The noninteracting Hamiltonian $\tilde H_z$ is
\begin{widetext}
\begin{equation}
   \tilde H_{z} =  \frac{1}{2}  \sum_{\substack{m,n,\\ \delta, \tau,\sigma, \sigma' }}  \int d x\ \Big( i \delta\tau \Delta_c
     \psi_{mn 1 \delta \tau \sigma}  \sigma_y^{\sigma \sigma'}   \psi_{mn\bar 1\delta\tau \sigma'}  
     -  i\beta \tau \psi^{\dagger}_{mn 1 \delta \tau \sigma}    \sigma_x^{\sigma \sigma'}  \psi_{mn
     \bar 1\delta \tau \sigma'}+\ \text{H.c.} \Big).\label{intracellHzAppendix1}
\end{equation}
Decomposing the field operators $\psi$ into left and right movers expressing them in terms of the pseudolayer and pseudo-spin indices $\kappa,\nu$ [see Eq. \eqref{eq:unitary_transformation}], for the interior modes, we find

\begin{align}
   \tilde H_{z} =  \frac{1}{2}  \sum_{\substack{m,n \\ \delta, \kappa}}  \int dx  \Big[  \delta \Delta_c (
     \bar L_{mn 1 \delta \kappa\kappa}   \bar R_{mn\bar 1\delta\kappa \bar \kappa} -\bar R_{mn 1 \delta \kappa\bar \kappa}    \bar L_{mn\bar 1\delta\kappa \kappa} ) 
     -  i\beta \kappa (\bar L^{\dagger}_{mn 1 \delta \kappa \kappa}      \bar R_{mn
     \bar 1\delta \kappa \bar \kappa} + \bar R^{\dagger}_{mn 1 \delta \kappa \bar \kappa}    \bar L_{mn
     \bar 1\delta \kappa \kappa}) + \mathrm{H.c.}   \Big]. \label{intracellHzAppendix2}
\end{align} 
Dressing the four terms introduced above, we obtain 
\begin{align}
   \tilde H'_{z} &=  \frac{1}{2}  \sum_{\substack{m,n,\\ \delta, \kappa}}  \int d x\ \Bigg[ 
   \delta \Delta_c'
   \left(\bar L_{mn 1 \delta \kappa\kappa} \bar R_{mn\bar 1\delta\kappa \kappa} \right)^{p} 
   \bar L_{mn 1 \delta \kappa\kappa} \bar R_{mn\bar 1\delta\kappa \bar \kappa} 
   \left( \bar L_{mn 1 \delta \kappa \bar \kappa} \bar R_{mn\bar 1\delta\kappa \bar \kappa}\right)^p 
   \nonumber\\
  & -\, i\beta' \kappa 
   \left(\bar L^{\dagger}_{mn 1 \delta \kappa \kappa} \bar R_{mn\bar 1\delta\kappa \kappa}\right)^{p} 
   \bar L^{\dagger}_{mn 1 \delta \kappa \kappa} \bar R_{mn\bar 1\delta\kappa \bar \kappa}
   \left(\bar L^{\dagger}_{mn 1 \delta \kappa \bar \kappa} \bar R_{mn\bar 1\delta\kappa \bar \kappa}\right)^{p}
   +\ \text{H.c.} \Bigg]
   \nonumber\\
   &+\ \Bigg[
   -\,\delta\Delta_c'
   \left(\bar R_{mn 1 \delta \kappa\bar \kappa} \bar L_{mn\bar 1\delta\kappa \bar \kappa}\right)^{p}
   \bar R_{mn 1 \delta \kappa\bar \kappa} \bar L_{mn\bar 1\delta\kappa \kappa}
   \left(\bar R_{mn 1 \delta \kappa\kappa} \bar L_{mn\bar 1\delta\kappa \kappa}\right)^{p}
   \nonumber\\
   & -\, i\beta' \kappa
   \left(\bar R^{\dagger}_{mn 1 \delta \kappa \bar \kappa} \bar L_{mn\bar 1\delta\kappa \kappa}\right)^{p}
   \bar R^{\dagger}_{mn 1 \delta \kappa \bar \kappa} \bar L_{mn\bar 1\delta\kappa \kappa}
   \left(\bar R^{\dagger}_{mn 1 \delta \kappa \kappa} \bar L_{mn\bar 1\delta\kappa \kappa}\right)^{p}
   +\ \text{H.c.} \Bigg]
   \label{intracellHzAppendixdressed}
\end{align}
such that $p=(q-1)/2.$
Upon bosonizing the fields $\bar L_{mn\gamma\delta\kappa\nu}$ and $\bar R_{mn\gamma\delta\kappa \nu}$ using Eq. \eqref{bosonizationbar} and then refermionizing them using Eq. \eqref{refermionization}, we find

\begin{align}
   \tilde H'_{z} 
   =  \frac{1}{2}  \sum_{\substack{m,n,\\ \delta, \kappa}}  \int d x\ \Big[ &\delta \Delta_c'
     \bar L^{(q)}_{mn 1 \delta \kappa\kappa}   \bar R^{(q)}_{mn\bar 1\delta\kappa \bar \kappa}  
     -  i\beta' \kappa (\bar L^{(q)}_{mn 1 \delta \kappa \kappa})^{\dagger}      \bar R^{(q)}_{mn
     \bar 1\delta \kappa \bar \kappa}+\ \text{H.c.} \Big] 
     \nonumber \\
     +
     \Big[ &-\delta\Delta_c'
     \bar R^{(q)}_{mn 1 \delta \kappa\bar \kappa}    \bar L^{(q)}_{mn\bar 1\delta\kappa \kappa} 
     -  i\beta' \kappa (\bar R^{(q)}_{mn 1 \delta \kappa \bar \kappa})^{\dagger}    \bar L^{(q)}_{mn
     \bar 1\delta \kappa \kappa}+\ \text{H.c.} \Big], \label{intracellHzAppendixdressedafterrefermionization}
\end{align} 
where $\bar L^{(q)}$ and $\bar R^{(q)}$ are chiral, composite fermion operators, as discussed in the main text. We can now rewrite Eq. \eqref{intracellHzAppendixdressedafterrefermionization} in terms of the parafermion operators $\chi^{(q)(R/L)}_{mn\gamma\delta\kappa\nu}$ and $\bar \chi^{(q)(R/L)}_{mn\gamma\delta\kappa\nu}$ to arrive at 
\begin{align}
    \tilde H_{z}' = \frac{i}{2} \sum_{m,n,\delta,\kappa} \int dx \, \Big[
      (\kappa\beta' - \Delta_c') 
\chi^{(q)\kappa R}_{mn \bar 1 \delta} \chi^{(q)\kappa L}_{mn1 \delta}
 +  (\kappa\beta' + \Delta_c') 
\chi^{(q)\kappa L}_{mn \bar 1 \delta} \chi^{(q)\kappa R}_{mn1 \delta} 
 \Big],
\end{align} 
\end{widetext}
which is the same as $\tilde H_z$ in Eq. \eqref{intra_Hz_as_Majorana} if we let $\chi^{\kappa (R/L)}_{mn\gamma\delta} \rightarrow\chi^{(q) \kappa (R/L)}_{mn\gamma\delta}$. Similarly, we can 
dress the terms $H_z$, $\tilde H_y$, $H_y$ to find the corresponding Hamiltonians $H'_z$, 
$\tilde H_y'$, $H'_y$. We provide the final expressions for these terms in Sec.~\ref{sec:Parafermion} of the main text, where we find that they are also a verbatim reproduction of their noninteracting counterparts up to the replacement $\chi^{\kappa (R/L)}_{mn\gamma\delta} \rightarrow\chi^{(q) \kappa (R/L)}_{mn\gamma\delta}$.  Thus, we may conclude that there is indeed a Kramers pair of counterpropagating $\mathbb Z_{2q}$ parafermionic hinge modes along a one-dimensional hinge of the system.

\bibliography{Manuscript/Citations}
        
\end{document}